\documentclass[5p,times]{elsarticle}
\usepackage{amsmath,amssymb,amsthm}
\usepackage{algorithm}
\usepackage{algorithmic}
\usepackage{graphicx}
\usepackage{booktabs}
\usepackage[section]{placeins}
\usepackage{multirow}
\usepackage[x11names,table]{xcolor}
\usepackage{tabularx}
\usepackage{gensymb}

\counterwithout{equation}{section}
\usepackage{makecell}
\usepackage{lineno}

\usepackage[colorlinks=true,           
            linkcolor=blue,            
            citecolor=blue,            
            urlcolor=blue]{hyperref}
\usepackage{xurl}

\usepackage[figurename=Fig. ]{caption}
\usepackage{subcaption}
\usepackage{wrapfig}
\usepackage[intoc]{nomencl}
\usepackage{etoolbox}
\usepackage{framed} 
\makenomenclature

\renewcommand{\nomgroup}[1]{%
  \item[\bfseries
    \ifstrequal{#1}{S}{Sets}{%
    \ifstrequal{#1}{I}{Indices}{%
    \ifstrequal{#1}{V}{Variables}{%
    \ifstrequal{#1}{P}{Parameters}{}}}}%
  ]
}

\theoremstyle{plain}

\theoremstyle{remark}

\begin{document}
\begin{frontmatter}

\title{Grid Capacity Expansion under Data Centers and Electrified Manufacturing Large Loads}
\author[inst1]{Jiyong Lee}
\author[inst4]{Melody Agustin}
\author[inst4]{Joanne Langsdorf}
\author[inst2]{Erhan Kutanoglu}
\author[inst1,inst3]{Michael Baldea}
\author[inst1]{Ilias Mitrai\corref{cor1}}
\ead{imitrai@che.utexas.edu}
\cortext[cor1]{Corresponding author}

\affiliation[inst1]{%
  organization={McKetta Department of Chemical Engineering, The University of Texas at Austin},
  city={Austin},
  postcode={TX 78712},
  country={United States}
}
\affiliation[inst2]{%
  organization={Walker Department of Mechanical Engineering, The University of Texas at Austin},
  city={Austin},
  postcode={TX 78712},
  country={United States}
}
\affiliation[inst3]{%
  organization={Oden Institute for Computational Engineering and Sciences, The University of Texas at Austin},
  city={Austin},
  postcode={TX 78712},
  country={United States}
}
\affiliation[inst4]{%
  organization={TotalEnergies Research \& Technology USA, LLC},
  city={Houston},
  postcode={TX 77002},
  country={United States}
}

\begin{abstract}

In this paper, we consider the expansion of power grids under emerging large loads from data centers and electrified manufacturing. We develop a multi-period grid capacity expansion model to determine optimal investment profiles for power generation, storage, and transmission capacity while accounting for hourly power dispatch, such that electricity demand is satisfied and the total planning and operation cost is minimized. We also propose a new modeling approach regarding the spatial distribution of demand from large loads. The model is used to analyze the expansion of a synthetic grid that follows key characteristics of the ERCOT system over a seven-year planning horizon, under loads from data centers and electrified oil refining, which account for 17.5\% and 4.7\% of total annual electricity demand by the end of the planning horizon. The optimal investment policy leads to an 83.6\% increase in generation capacity and exploits the short construction times of solar and storage as well as the operational flexibility of thermal generators. Finally, sensitivity analysis reveals that the construction time of grid assets substantially impacts investment timing, generation technology mix, and transmission capacity expansion. The proposed modeling framework is general and can be extended to other grid systems, enabling the exploration of diverse demand scenarios, policy assumptions, and regional characteristics.

\end{abstract}


\end{frontmatter}

\section{Introduction}\label{sec:intro}

Large loads are reshaping the operation and planning of power systems by altering the magnitude, temporal dynamics, and spatial distribution of electricity demand. Historically, in the United States, electricity demand has grown at an annual rate of approximately $2\%$ \cite{EIA_MER_2025}. However, recent forecasts project a significantly higher annual demand increase rate across the US. For example, the Electric Reliability Council of Texas (ERCOT) is projected to experience significant demand growth within a decade \cite{ERCOT2025LTLF}. For the past six years (2019 to 2025) ERCOT's annual electricity demand has grown at approximately a 4.0\% compound annual growth rate, and for the following six years from 2025 to 2031, it is projected to grow at 13.5\% annually \cite{ERCOT2025LTLF}. 

 
A significant portion of this growth can be attributed to data center and electrified manufacturing. Data centers are projected to account for 6.7\% to 12\% of the total U.S. electricity consumption in 2028 \cite{shehabi20242024}, which is equivalent to 326 to 578 TWh. In parallel, electrified manufacturing represents a substantial source of new electricity demand. In 2018, on-site process heating across U.S. manufacturing consumed 1,466 TWh \cite{DOE_ITO_2018_MECS}. Assuming a 97\% Joule-heating efficiency, the electrification of the entire manufacturing sector would require approximately 173 GW of power only for process heating. The simultaneous growth of these loads will require significant investments in grid capacity expansion.

Grid capacity expansion planning is a well-studied problem that considers decisions at both yearly (i.e., investment in new capacity) and hourly (e.g., power dispatch) levels \cite{hemmati2013state, koltsaklis2018state}. These models have been used to identify grid expansion driven by population-based demand growth, decarbonization goals, or public policies. A major challenge is computational tractability, since the model complexity increases significantly with the planning horizon and grid size. Two approaches are usually followed to address this issue. The first relies on time clustering to identify a few representative days that capture key features of the demand and renewable availability \cite{nahmmacher2016carpe, liu2017hierarchical, li2022representative}. The alternative is to study simplified systems with few nodes (buses) and transmission lines\cite{lara2018deterministic, potts2025grid, li2022mixed}. 

Although such simplifications lead to computationally tractable problems, the insights obtained can be misleading since the grid resolution is not adequate to identify issues related to congestion and power generation \cite{frew2016temporal, frysztacki2021strong, aryanpur2021review, serpe2025importance, glista2025zonal}. Moreover, this aggregated spatial system representation does not allow the detailed modeling of the spatial distribution of large loads. Finally, current models for grid expansion usually consider large loads from a single sector, such as data centers \cite{potts2025grid, shao2025stochastic, abdelhady2026optimal}, electrolyzers  \cite{lieberwirth2023decarbonizing}, and electric heating \cite{gailani2025assessing}. However, the effect of the simultaneous demand growth from data centers and manufacturing has not been studied. Although both sectors will create new demand in the form of large loads, they differ in key ways that complicate grid expansion.

Data centers have shorter construction times compared to electricity generation and transmission assets, creating a mismatch between demand and capacity growth. Moreover, data center siting is influenced by the availability of reliable electricity supply, adequate water resources for cooling, and a skilled workforce \cite{arzumanyan2025geospatial}. As a result, data centers are expected to cluster geographically in major urban areas \cite{arzumanyan2025geospatial}, creating localized grid overload and congestion. From an operational perspective, power consumption is mostly tied to powering and cooling the computing infrastructure. Power Usage Effectiveness (PUE) is a key metric that indicates how much overhead energy is required to operate data centers. It is defined as the ratio of the total electricity consumption of a data center to the electricity consumption of its IT equipment. According to the 2024 U.S. data center energy usage report \cite{shehabi20242024}, the annual average PUE in 2023 was 1.4, implying that approximately 71\% of total data center electricity is used for IT operations, while the remaining 29\% is consumed by supporting systems such as cooling and other infrastructure. Data centers can potentially adjust energy consumption either via demand response mechanisms \cite{kwon2018demand, ding2018emission, bahrami2018data, yu2016distributed} or computing task reallocation \cite{hung2015scheduling, hu2017time, dogar2014decentralized}. In general, data centers either provide basic services, such as storage, and have relatively steady power demand throughout the day, or are used for training AI models, in which case their power consumption can adjust within seconds. The current trend is towards the construction of hyperscale data centers with projected power demand on the order of hundreds of megawatts or even gigawatts \cite{shehabi20242024, NREL_SpeedToPower}.

Industrial electrification is emerging as a key pathway to improve the efficiency of energy-intensive industries and reduce their carbon intensity \cite{baldea2025transforming}. Recently, process heating has been considered a candidate for electrification, where substituting traditional fossil-based process heating with electric heating can improve energy efficiency and reduce Scope 1 emissions \cite{giannikopoulos2024thermal, chattopadhyay2025optimal, rho2025probing, granacher2025system}. However, connecting such manufacturing systems to the grid presents new challenges. Unlike data centers, the deployment of electrified manufacturing technologies will occur over a longer time horizon due to the longer construction times and large capital costs associated with retrofitting and/or building new facilities. Moreover, their production is often relatively stable, although variability can arise depending on operational conditions. Even if certain manufacturing systems can provide grid flexibility through demand response \cite{zhang2016enterprise, tang2023grid}, the electrification of the entire manufacturing industry will impose a significant load on the grid due to the magnitude of demand, its relatively constant profile, and the geographical clustering of certain manufacturing hubs.



In this regard, this work seeks to address the following research questions:
\begin{enumerate}
    \item What is the economically optimal grid capacity expansion and operation strategy to accommodate the abrupt demand growth from data centers and electrified manufacturing over planning horizons?
    \item How does the temporal mismatch between demand realization and construction times of grid assets shape investment timing, technology selection, and overall system costs under rapid large load growth?
\end{enumerate}

Motivated by the above, in this paper, we develop a multi-period grid capacity expansion model to analyze the joint effect of data center and electrified manufacturing growth on grid capacity expansion. The model considers strategic investments, such as expansion of generation facilities, and operational decisions, i.e., hourly dispatch. The model accounts for both temporal and spatial evolution of large loads across the entire planning horizon. The resulting model is linear, which improves computational tractability when analyzing realistic grid systems with a large number of buses and long planning horizons. This is particularly important when considering different types of large loads. For example, in Texas, data centers tend to be located near major cities, while oil extraction activities occur mostly in the west and processing and refining in the coastal region. This spatial heterogeneity requires a detailed network representation with sufficient geospatial resolution to identify the location-specific (i.e., nodal) optimal generation and transmission investments. 

We use the proposed model to analyze the expansion of a synthetic grid system that reflects generation and transmission features of ERCOT and must satisfy demand from data centers and electrified oil refining. We use ERCOT's projection about base and data center demand and assume that $30\%$ of the entire process heat demand from oil refining in Texas is electrified until year 2031 with $5\%$ annual increment. The hourly data center and electrified manufacturing load is assumed to be constant within each year. Under these assumptions, electricity demand from data centers and electrified oil refineries accounts for 17.5\% and 4.7\% of total electricity demand by 2031, respectively. 

The results show that the total grid generation capacity should increase by approximately 83.6\% over seven years. The optimal investment strategy is largely driven by natural gas and solar generation, supported by battery storage, while transmission expansion is fairly limited. Solar generation is deployed at the beginning of the planning horizon because of its shortest construction lead time, thereby reducing demand curtailment. On the other hand, natural gas provides reliable and flexible generation to satisfy time-varying base loads and the inflexible large loads. Sensitivity analyses show that construction time assumptions significantly affect both timing and the technology mix, and that longer construction times for natural gas lead to investments in wind and solar generation as well as transmission expansion. These results show that construction time is a critical factor in grid capacity expansion planning under large load growth and a short planning horizon.

The rest of the paper is organized as follows. In Section~\ref{sec: math model} we present the mathematical formulation of the grid capacity expansion planning model and in Section~\ref{sec: modeling load} we present the modeling of both base and large loads. In Section~\ref{sec: case study}, we describe the case study, in Section~\ref{sec: results} we discuss the results, and in Section~\ref{sec: concl}, we conclude the work and outline directions for future research.

\section{Multi-period grid capacity expansion planning model}\label{sec: math model}
In this section, we present the multi-period capacity expansion planning model. We assume a power grid with $N$ buses and $N_{\rm{I}}$ generation technologies, including both thermal $(N_{\rm{TH}})$ and renewable generators $(N_{\rm{RN}})$, and each bus can have multiple generation technologies. The grid must satisfy two types of demand: base demand from residential, industrial, and transportation sectors, and emerging large loads from data centers and electrified manufacturing. Given this emerging demand, we seek to find the optimal expansion of existing generation, transmission, and storage assets over a planning horizon of $N_{\rm{T}}$ years. We consider the scenario in which capacity can be expanded only for existing generation and transmission assets; we do not consider adding new buses, new technologies per bus, or transmission lines to the grid. This assumption is made to improve tractability and the model can be readily adapted to account for new buses, generator technologies, and transmission lines.

We define $\mathcal{T} = \{1,\dots,N_{\rm{T}}\}$ as the set of planning years, $\mathcal{N} = \{1,\dots, N\}$ as the set of buses in the grid, $\mathcal{I}=\{1,\dots, N_{\rm{I}}\}$ as the set of all generation technologies, $\mathcal{I}_{\rm{TH}}=\{1,\dots, N_{\rm{TH}}\}$ as the set of thermal generators, and $\mathcal{I}_{\rm{RN}} = \{1,\dots,N_{\rm{RN}}\}$ as the set of renewable generators. We also define $N_{\rm{d}}$ $(\mathcal{D}=\{1,\dots, N_{\rm{d}}\})$ as the number of representative days to capture the daily and hourly temporal evolution of the base demand and power grid operation. Finally, we define $\mathcal{H}$ as the set of hours in a day. 

We assume that the locations of data centers and electrified manufacturing facilities are known at the regional level, and we define $N_{\rm{c}}$ as the number of geographical regions containing data centers and $N_{{e}}$ as the number of regions with electrified manufacturing. We define the corresponding sets for each set of regions as $\mathcal{C} = \{1,\dots, N_{\rm{c}}\}$ and $\mathcal{E} = \{1,\dots, N_{\rm{e}}\}$, respectively. Depending on the available information about the location of the large loads, a region can correspond to a bus, a country, or a weather region.

\subsection{Capacity expansion constraints}
We define variable $c^{\rm{gen}}_{nit}$ as the new generation capacity at bus $n$ for generation technology $i$ in year $t$, $c^{\rm{trans}}_{nn't}$ as the new transmission capacity between buses $n$ and $n'$ in year $t$, and $c^{\rm{stor}}_{nt}$ as the new storage capacity at bus $n$ in year $t$. These investment decisions are bounded as follows
\begin{equation}
\begin{aligned}
& 0 \leq c^{\rm{gen}}_{nit} \leq \bar{c}^{\rm{gen}} \ && \forall n \in \mathcal{N}, i \in \mathcal{I}(n), t \in \mathcal{T}\\
& 0 \leq c^{\rm{trans}}_{nn't} \leq \bar{c}^{\rm{trans}} \ && \forall n \in  \mathcal{N}, n' \in \mathcal{N}(n), t \in \mathcal{T}\\
& 0 \leq c^{\rm{stor}}_{nt} \leq \bar{c}^{\rm{stor}} \ && \forall n \in \mathcal{N}, t \in \mathcal{T},
\end{aligned}
\end{equation}
where $\bar{c}^{gen}$, $\bar{c}^{\rm{trans}}$, $\bar{c}^{\rm{stor}}$ are the maximum capacities for the new generators, new transmission lines, and new storage, respectively. $\mathcal{I}(n)$ denotes the set of generation technologies available at bus $n$ and $\mathcal{N}(n)$ denotes the set of neighboring buses for bus $n$. For each year $t$ the available capacity for generation, transmission, and storage is equal to
\begin{equation}
\begin{aligned}
    C^{\rm{gen}}_{nit} & = c^{\rm{gen},0}_{ni}+\sum_{t'=1}^{t-\omega_{i}^{\rm{gen}}} c^{\rm{gen}}_{nit'} \ && \forall n\in \mathcal{N}, i \in \mathcal{I}, t \in \mathcal{T}\\
    C^{\rm{trans}}_{nn't} & = c^{\rm{trans},0}_{nn'}+\sum_{t'=1}^{t-\omega^{\rm{trans}}} c^{\rm{trans}}_{nn't'} \ && \forall n\in \mathcal{N}, n' \in \mathcal{N}(n), n<n', t \in \mathcal{T}\\
    C^{\rm{stor}}_{nt} & = c^{\rm{stor},0}_{n} + \sum_{t'=1}^{t-\omega^{\rm{stor}}} c^{\rm{stor}}_{nt'} \ && \forall n \in \mathcal{N}, t \in \mathcal{T},  
\end{aligned}
\end{equation}
where $c^{\rm{gen},0}_{nit}$, $c^{\rm{trans}}_{nn'}$, $c^{\rm{stor},0}_{n}$ are the initial generation, storage, and transmission capacities, $\omega^{\rm{gen}}_{i}$ is the construction time of generation technology $i$, $\omega^{\rm{trans}}$ is the construction time for transmission lines, and $\omega^{\rm{stor}}$ is the construction time for storage.
These constraints capture the effect of construction time on the availability of new capacity. For example, a new investment in generation technology $i$ at bus $n$ and year $t$ will become available at year $t+\omega^{\rm{gen}}_{i}$. 

The system-wide total generation capacity must be equal to or greater than the annual peak load of the grid each year, which is defined as $P^{\rm{peak}}_{t}$. This is enforced via the following constraints
\begin{equation}
P^{\rm{peak}}_{t}\leq\sum_{n\in\mathcal{N}}\sum_{i\in \mathcal{I}(n)} C^{\rm{gen}}_{nit} \ \  \forall t\in \mathcal{T}.
\label{eq: peakLoad}
\end{equation}

\subsection{Grid operation constraints}
The constraints presented above capture the evolution of the grid capacity with time. In this section, we present the constraints related to the operation of the grid. First, we focus on power conservation at each bus. We define variable $p^{\rm{gen}}_{nitdh} \in \mathbb{R}_{+}$ as the power generated at bus $n$ from generation technology $i$ in year $t$ day $d$ and hour $h$, and $p^{\rm{curt,gen}}_{nitdh} \in \mathbb{R}_{+}$ as the curtailment of generation from technology $i$ at bus $n$ in year $t$ day $d$ and hour $h$, and $p^{\rm{curt,dem}}_{ntdh} \in \mathbb{R}_{+}$ as the curtailment of demand at bus $n$ day $d$ hour $h$ and year $t$. We also define $p^{\rm{charge}}_{ntdh} \in \mathbb{R}_{+}$ and $p^{\rm{disch}}_{ntdh}\in \mathbb{R}_{+}$ as the power for charging and discharging the battery storage at each bus $n$ in year $t$ day $d$ and hour $h$. Using the DC-Optimal Power Flow (DC-OPF) approximation, the power balance at bus $n$, hour $h$, day $d$, and year $t$ is
\begin{equation}
\label{eq: powerBalance}
\begin{aligned}
& \sum_{i \in \mathcal{I}(n)} \left(p^{\rm{gen}}_{nitdh} - p^{\rm{curt,gen}}_{nitdh}\right) 
-\sum_{n' \in \mathcal{N}(n)} p^{\rm{trans}}_{nn'tdh} + p^{\rm{disch}}_{ntdh} - p^{\rm{charge}}_{ntdh}\\
 & = D^{\rm{base}}_{ntdh} + p^{\rm{DC}}_{ntdh} + p^{\rm{EOR}}_{ntdh} - p^{\rm{curt,dem}}_{ntdh} 
\end{aligned}
\end{equation}
where $D^{\rm{base}}_{ntdh}$ is the base load at bus $n$ year $t$ day $d$ and hour $h$, which will be discussed in detail in Section~\ref{sec: modeling load}. $p^{\rm{DC}}_{ntdh}\in\mathbb{R_{+}}$ is the power used to satisfy data center demand at bus $n$ year $t$ and hour $h$, and $p^{\rm{EOR}}_{ntdh}\in \mathbb{R}_{+}$ is the power used to satisfy demand from electrified manufacturing. The power transmitted between bus $n$ and $n'$ for year $t$ day $d$ and hour $h$ is computed via the following constraint
\begin{equation}
\label{eq: transPower}
p^{\rm{trans}}_{nn'tdh}=\frac{S_{\rm{base}}}{X_{nn'}}(\theta_{ntdh}-\theta_{n'tdh}),
\end{equation}
where $\theta_{ntdh} \in [-\pi,\pi]$ is the voltage angle in radians. We assume that the transmission lines are bi-directional, the system based power $S_{\rm{base}}$ is 100 MVA, and $X_{nn'}$ is the line reactance.

The load from data centers and electrified manufacturing is satisfied at the regional level, i.e., multiple buses contribute to satisfy the demand. We define $D^{\rm{DC}}_{ctdh}$ as the load from data center for region $c$, year $t$, day $d$, and hour $h$, and $D^{\rm{EOR}}_{etdh}$ as the load from electrified manufacturing at region $e$, year $t$, day $d$, and hour $h$. We note that the geographic regions for the data centers and electrified manufacturing facilities are not necessarily the same. The region-level load satisfaction for the large loads is enforced via the following constraints
\begin{equation}
    \sum_{n \in \mathcal{N}_{c}} p^{\rm{DC}}_{ntdh} = D^{\rm{DC}}_{ctdh} \quad \forall c \in \mathcal{C},t\in \mathcal{T}, d \in \mathcal{D}, h \in \mathcal{H},
    \label{eq: county-level DC load satisfaction}
\end{equation}
\begin{equation}
    \sum_{n \in \mathcal{N}_{e}} p^{\rm{EOR}}_{ntdh} = D^{\rm{EOR}}_{etdh} \quad \forall e \in {E}, t\in \mathcal{T}, d \in \mathcal{D}, h \in \mathcal{H},
    \label{eq: county-level EOR load satisfaction}
\end{equation}
where $\mathcal{N}_{c}$ is the set of buses in region $c$ and $\mathcal{N}_{e}$ is the set of buses in region $e$. These load satisfaction constraints couple the spatial scale where large loads are located, i.e., counties, with the grid-level scale, i.e., buses. Moreover, these constraints allow the reallocation of the large loads within the buses in each region. As described above, in this modeling approach, we assume that the large load demand is known at a regional level. If the demand is known at a bus level, either for some or all buses, then this can be incorporated directly in the power balance constraint (Eq.~\ref{eq: powerBalance}) as an additional demand source.

The available generation capacity at each bus constrains the power that can be generated. For the thermal generators, this is  enforced via the following constraints
\begin{equation}
\begin{aligned}
    & p^{\rm{gen}}_{nitdh} \geq F^{\rm{min}}_{i} C^{\rm{gen}}_{nit} \ \ \forall n\in \mathcal{N}, i \in \mathcal{I}_{TH} (n), t \in \mathcal{T}, d \in \mathcal{D}, h \in \mathcal{H},
    \\ & p^{\rm{gen}}_{nitdh} \leq F^{\rm{max}}_{i} C^{\rm{gen}}_{nit}\ \ \forall n\in \mathcal{N}, i \in \mathcal{I}_{TH} (n), t \in \mathcal{T}, d \in \mathcal{D}, h \in \mathcal{H},
\end{aligned}
\label{eq: gen_CF_TH}
\end{equation}
where $F^{\rm{min}}_{i}$ and $F^{\rm{max}}_{i}$ are the minimum and maximum capacity factors of thermal generators. These equations assume that thermal generators can adjust their power output within the limits imposed by capacity factors and that the generators are always operational. Although generators can be turned on and off based on demand, we assume they are always used to avoid introducing binary variables. 

The power generated by renewable generators, i.e., solar, wind, and hydro, is constrained by the available capacity and the capacity factor at each but $n$, hour $t$, day $d$, and year $t$. This is enforced as follows
\begin{equation}
\label{eq: capacityFactor_renewable}
\begin{aligned}
    &p^{\rm{gen}}_{nitdh} = F^{\rm{RN}}_{nidh} C^{\rm{gen}}_{nit}
     \ \ \forall n\in \mathcal{N}, i \in \mathcal{I}_{\rm{RN}}(n), t\in \mathcal{T}, d\in \mathcal{D}, h \in \mathcal{H},
\end{aligned}
\end{equation}
where $F^{\rm{RN}}_{nidh}$ is the capacity factor of the renewable generator $i$ at bus $n$, day $d$, and hour $h$. This capacity factor depends on weather and geological conditions, and is assumed to exhibit identical trends each planning year.

The change in the power generated by each thermal generator within an hour is limited by its ramping rate $R^{\rm{ramp}}_{i}$, as follows
\begin{equation}
\begin{aligned}
     &p^{\rm{gen}}_{nitdh} - p^{\rm{gen}}_{nitd,h-1} \geq - R^{\rm{ramp}}_{i} C^{\rm{gen}}_{nit}  \\ & \ \forall n\in \mathcal{N}, i \in \mathcal{I}_{\rm{TH}}(n),t\in \mathcal{T}, d\in \mathcal{D}, h \in \mathcal{H}\setminus \{1\} \\
     &p^{\rm{gen}}_{nitdh} - p^{\rm{gen}}_{nitd,h-1} \leq R^{\rm{ramp}}_{i} C^{\rm{gen}}_{nit} \\ & \ \forall n\in \mathcal{N}, i \in \mathcal{I}_{\rm{TH}}(n),t\in \mathcal{T}, d\in \mathcal{D}, h \in \mathcal{H}\setminus\{1\}
\label{eq: rampingRate_thermalGen}
\end{aligned}
\end{equation}
\begin{equation}
\begin{aligned}
     &p^{\rm{gen}}_{nitd,1} - p^{\rm{gen}}_{nit,d-1,24} \geq - R^{\rm{ramp}}_{i} C^{\rm{gen}}_{nit} \\ & \ \forall n\in \mathcal{N}, i \in \mathcal{I}_{\rm{TH}}(n),t\in \mathcal{T}, d\in \mathcal{D}\setminus \{1\}
     \\ &p^{\rm{gen}}_{nitd,1} - p^{\rm{gen}}_{nit,d-1,24} \leq R^{\rm{ramp}}_{i} C^{\rm{gen}}_{nit} \\ & \ \forall n\in \mathcal{N}, i \in \mathcal{I}_{\rm{TH}}(n),t\in \mathcal{T}, d\in \mathcal{D}\setminus \{1\},
\end{aligned}
\label{eq: rampingRate_thermalGen_dayConnection}
\end{equation}
Eq.~\ref{eq: rampingRate_thermalGen} limits the rate of change between two consecutive hours, while Eq.~\ref{eq: rampingRate_thermalGen_dayConnection} limits the rate of change between the start of day $d$ and the end of day $d-1$ except the first day.

The power transmitted through a transmission line connecting bus $n$ and $n'$ on day $d$ and hour $h$ is constrained by the transmission line capacity as follows
\begin{equation}
\begin{split}
&p^{\rm{trans}}_{nn'tdh} \geq -C^{\rm{trans}}_{nn't} \ \ \forall n \in \mathcal{N}, n' \in \mathcal{N}(n), n < n', t\in \mathcal{T}, d \in \mathcal{D}, h \in \mathcal{H}
\\ &p^{\rm{trans}}_{nn'tdh} \leq   C^{\rm{trans}}_{nn't} \ \ \forall n \in \mathcal{N}, n' \in \mathcal{N}(n), n < n', t\in \mathcal{T}, d \in \mathcal{D}, h \in \mathcal{H}.
\end{split}
\end{equation}

Finally, we consider the dynamics and operation of the storage. We define $e^{\rm{stor}}_{ntdh}$ as the energy level of the storage at bus $n$ in year $t$, day $d$, and hour $h$. The level of energy stored depends on the amount of energy deposited from the grid, $p^{\rm{charge}}_{ntdh}$, and discharged $p^{\rm{disch}}_{ntdh}$. The operation of the storage is captured via the following constraints
\begin{equation}
\label{eq: stor_energyLevel}
\begin{aligned}
    &e^{\rm{stor}}_{ntdh} = e^{\rm{stor}}_{ntdh-1} + \eta^{\rm{charge}} p^{\rm{charge}}_{ntdh}- \frac{1}{\eta^{\rm{disch}}} p^{\rm{disch}}_{ntdh}
    \\ & \ \forall n\in \mathcal{N},t\in \mathcal{T}, d \in \mathcal{D}, h\in \mathcal{H}
\end{aligned} 
\end{equation}
\begin{equation}
\label{eq: stor_energyLevel_dayConnection}
\begin{aligned}
    &e^{\rm{stor}}_{ntd1} = e^{\rm{stor}}_{n,t,d-1,24}
     \ \ \forall n\in \mathcal{N},t\in \mathcal{T}, d\in \mathcal{D}\setminus \{1\}
\end{aligned} 
\end{equation}
\begin{equation}
\label{eq: stor_capacity}
\begin{aligned}
   &e^{\rm{stor}}_{ntdh} \leq C^{\rm{stor}}_{nt} H^{\rm{stor}}
    \ \ \forall n\in \mathcal{N},t\in \mathcal{T},\ d \in \mathcal{D},\ h\in \mathcal{H}
\end{aligned} 
\end{equation}
\begin{equation}
\label{eq: stor_chargeCapacity}
\begin{aligned}
    &p^{\rm{charge}}_{ntdh}\leq C^{\rm{stor}}_{nt} 
    \ \ \forall n\in \mathcal{N},t\in \mathcal{T},\ d \in \mathcal{D},\ h\in \mathcal{H}
\end{aligned} 
\end{equation}
\begin{equation}
\label{eq: stor_dischargeCapacity}
\begin{aligned}
    &p^{\rm{disch}}_{ntdh}\leq C^{\rm{stor}}_{nt}
    \ \ \forall n\in \mathcal{N},t\in \mathcal{T},\ d \in \mathcal{D},\ h\in \mathcal{H},
\end{aligned} 
\end{equation}
where $\eta^{\rm{charge}}$ and $\eta^{\rm{disch}}$ are the charging and discharging efficiency factors, respectively, and $H^{\rm{stor}}$ is the storage duration in hours. The constraint (Eq.~\ref{eq: stor_energyLevel}) tracks the level of the stored energy based on the amount of energy charged $(p^{\rm{charge}}_{ntdh})$ from the grid and the amount of energy discharged $(p^{\rm{disch}}_{ntdh})$. The second constraint (Eq.~\ref{eq: stor_energyLevel_dayConnection}) ensures that the amount of energy stored at the beginning of day $d$ is the same as that of the end of day $d-1$. The third constraint (Eq.~\ref{eq: stor_capacity}) limits the maximum capacity of the stored energy level based on the available capacity. The fourth (Eq.~\ref{eq: stor_chargeCapacity}) and fifth constraints (Eq.~\ref{eq: stor_dischargeCapacity}) limit the maximum power charged and discharged from the battery based on the installed power capacity. We note that the unit of $e^{\rm{stor}}_{ntdh}$ is $MWh$, while the unit of $C^{\rm{stor}}_{nt}$ is $MW$.

\subsection{Objective function}\label{sec: obj}

The objective is the summation of the capital and operating costs over the entire planning horizon. The capital cost for year $t$, $\Phi^{\rm{CAPEX}}_{t}$, accounts for the capital cost of new generation, transmission, and storage capacity, and is equal to
\begin{equation}
\begin{split}
    \Phi^{\rm{CAPEX}}_{t} =\sum_{\substack{n \in \mathcal{N}\\ i \in \mathcal{I}(n)}} \alpha_{it}^{\rm{gen}}\ c^{\rm{gen}}_{nit} + \sum_{(n,n') \in \mathcal{N}'} \alpha^{\rm{trans}}_{t}\ c^{\rm{trans}}_{nn't}\ L_{nn'} 
    + \sum_{n \in \mathcal{N}} \alpha^{\rm{stor}}_{t}\ c^{\rm{stor}}_{nt},
\end{split}
\end{equation}
where $\mathcal{N}'=\{n \in \mathcal{N}, n' \in \mathcal{N}(n), n<n'\}$, and $\alpha^{\rm{gen}}_{it}$, $\alpha^{\rm{stor}}_{t}$, and $\alpha^{\rm{trans}}_{t}$ are the unit capital costs for generation technology $i$, storage, and transmission, respectively, in year $t$. $L_{nn'}$ is the distance of the transmission line connecting bus $n$ and $n'$ in miles.

The operating cost, $\Phi^{\rm{OPEX}}_{t}$, accounts for the fixed and variable costs for each technology and the curtailment costs from generation and demand. The operating cost for each year is computed as follows
\begin{equation}
\begin{split}
    & \Phi^{\rm{OPEX}}_{t}
    =  \sum_{n \in \mathcal{N}} \bigg(\sum_{i \in \mathcal{I}(n)} \beta^{\rm{gen}}_{it} C^{\rm{gen}}_{nit} + \beta^{\rm{stor}}_{t} C^{\rm{stor}}_{nt}\bigg) 
    \\+&\sum_{n \in \mathcal{N}} \sum_{d \in \mathcal{D}} w_{d} \sum_{h \in \mathcal{H}} \bigg(\sum_{i \in \mathcal{I}_{\rm{TH}}(n)} \Big(\hat{\gamma}_{it} p^{\rm{gen}}_{nitdh}
    + \zeta \ p^{\rm{curt,gen}}_{nitdh} \Big) + \delta \ p^{\rm{curt,dem}}_{ntdh} \bigg),
\end{split}
\end{equation}
where $\beta^{\rm{gen}}_{it}$ is the fixed operation and maintenance (FOM) cost for generation technology $i$ in year $t$, and $\beta^{\rm{stor}}_{t}$ is the FOM cost for storage in year $t$. $w_{d}$ is the weight of representative day $d$ capturing the fraction of days within a year represented by day $d$. $\hat{\gamma}_{it} = \gamma^{\rm{gen}}_{it} + \gamma^{\rm{fuel}}_{it} HR_{it}$ with $\gamma^{\rm{gen}}_{it}$ being the variable operation and maintenance (VOM) cost of thermal generation technology $i$ in year $t$, $\gamma^{\rm{fuel}}_{it}$ is the fuel cost of thermal generation technology $i$ in year $t$, and $HR_{it}$ is the heat rate of thermal generation technology $i$ in year $t$. $\zeta$, $\delta$ are the curtailment costs for generation and demand, respectively. For renewable generation assets (i.e., solar, wind, and hydro) and transmission, the variable operation and maintenance costs are assumed to be negligible.

The total cost over the entire planning horizon is equal to
\begin{equation}
\label{eq: total costs}
        \sum_{t \in \mathcal{T}} \frac{1}{(1+Ir)^{t-t_{\rm{base}}}}(\Phi^{\rm{CAPEX}}_{t}+\Phi^{\rm{OPEX}}_{t}),
\end{equation}
where $Ir$ is the interest rate, $t_{\rm{base}}$ is the financial base year. Note that all monetary values are depreciated to their net present value relative to the financial base year.

\section{Modeling large loads from data centers and electrified manufacturing}\label{sec: modeling load}

In this section, we consider the spatiotemporal modeling of the demand from data centers and electrified manufacturing. Unlike base demand from residential, industrial and transportation user base, for which historical data are available, modeling demand from data centers and manufacturing is more challenging because the locations of these future facilities and where and how they will be connected to the grid are not yet known.
We present a generic framework for modeling such demands and assume that the projection of total electricity demand over the planning horizon $E_{t}$ is known.

\subsection{Base load}
The base demand accounts for the aggregated demand from residential, industrial, and transportation sectors. We define $D^{\rm{base},0}_{ndh}$ as the initial base load profile at bus $n$ day $d$ and hour $h$, $E^{\rm{base}}_{t_{0}}$ as the original base electricity demand, and $E^{\rm{base}}_{t}$ as the projected annual base electricity demand for year $t$. We assume that the additional demand each year is distributed uniformly across buses and over all days and hours. Under this assumption, the base load for bus $n$ in year $t$, day $d$, and hour $h$ is computed as follows
\begin{equation}
\begin{aligned}
D^{\rm{base}}_{ntdh} = D^{\rm{base},0}_{ntdh} + \frac{E^{\rm{base}}_{t}-E^{\rm{base}}_{t_{0}}}{|\mathcal{N}||\mathcal{D}|| \mathcal{H}|}.
\end{aligned}
\label{eq: demand base}
\end{equation}

\subsection{Data center load}
The modeling of data center load presents two challenges. First, the exact locations of new data centers are not yet known. Moreover, data centers exhibit heterogeneous operational characteristics depending on their type. For example, data servers that provide basic services, such as data storage, operate at steady-state. On the contrary, data centers used for training artificial intelligence models can vary their power consumption greatly within a short amount of time \cite{shehabi20242024}. In this work, we assume that data center power consumption is satisfied by all the buses in a county, is constant over all operating hours within a given year, and changes across years according to the projected annual demand growth. Under this assumption, the large load demand at a county for a given year is constant; however, the amount of the demand satisfied by each bus in a county can vary with time. We define $P^{\rm{DC}}_{t}$ as the annual peak load of data center demand. The hourly data center load is then calculated using a load factor $LF^{\rm{DC}}$, assumed to be  $90\%$ \cite{CEC2024_DataCenterForecasts, Sherwood2025_LargeLoadTariffs}, i.e., the average data center load is equal to $P^{\rm{DC}}_{t} LF^{\rm{DC}}$. Overall, the data center demand at county $c$ year $t$, day $d$, and hour $h$ is equal to
\begin{equation}
\begin{aligned}
D^{\rm{DC}}_{ctdh} = \psi_{c} \frac{E^{\rm{DC}}_{t}}{|\mathcal{D}|| \mathcal{H}|} = \psi_{c} \frac{P^{\rm{DC}}_{t}\ LF^{\rm{DC}}\ 8760}{|\mathcal{D}|| \mathcal{H}|}. 
\end{aligned}
\label{eq: demand DC}
\end{equation}

\subsection{Electrified manufacturing load}
Similar to data centers, the modeling of the demand arising from electrified manufacturing also varies with time and space. We define $\phi_{t}$ as the electrification ratio in year $t$, i.e., the ratio of process heat generated from electrified heating technologies. This parameter determines the total electricity demand from the manufacturing sector each year, which equals $Q^{\rm{M}} \phi_{t} /\eta_{\rm{elec}}$, where $Q^{\rm{M}}$ is the total heat demand of the manufacturing sector, and $\eta_{\rm{elec}}$ is the Joule-heating efficiency for electrified technologies. In this work, we do not consider demand response or demand bidding of manufacturing process. Therefore, we assume that the demand from electrified manufacturing varies annually (due to $\phi_{t}$) but not daily or hourly, i.e., within each year, the demand is distributed equally across all days and hours. Similar to data centers, the demand from electrified manufacturing is constant each hour in a year at a county level, i.e., the amount of power generated by a bus and used to satisfy large loads from electrified manufacturing can vary with time.

To allocate this load spatially across the grid, we define $\psi_{e}$ as the ratio of heat duty of the manufacturing sector in geographic region $e$. This parameter together with the electrification ratio $\phi_{t}$, determines the magnitude of load in each region. We note that $\psi_{e}$ is not changing over time because manufacturing facilities have long lifetimes and construction times. The load from electrified manufacturing for each geographic region $e$, year $t$, day $d$, and hour $h$ is computed as follows
\begin{equation}
\begin{aligned}
D^{\rm{EM}}_{etdh} = \psi_{e} \frac{E^{\rm{EM}}_{t}}{|\mathcal{D}|| \mathcal{H}|} = \psi_{e} \frac{Q^{\rm{M}} \ (\phi_{t}/\eta_{\rm{elec}})\ 8760}{|\mathcal{D}|| \mathcal{H}|}.
\end{aligned}
\label{eq: demand EOR}
\end{equation}

\section{Case study}\label{sec: case study}
In this section, we use the proposed multi-period capacity expansion model to analyze the expansion of a synthetic grid from the literature \cite{lu2025synthetic} to accommodate large loads from data centers and electrified oil refining. We assume that the planing horizon is seven years from 2025 to 2031 ($N_{\rm{T}}=7)$. All input parameters and data sources used in the case study are summarized in Table~\ref{tab:input_parameters} for clarity.

\subsection{Reference grid} \label{sec: referenceGrid}
\begin{figure}[h]
    \centering
    \includegraphics[scale=0.4]{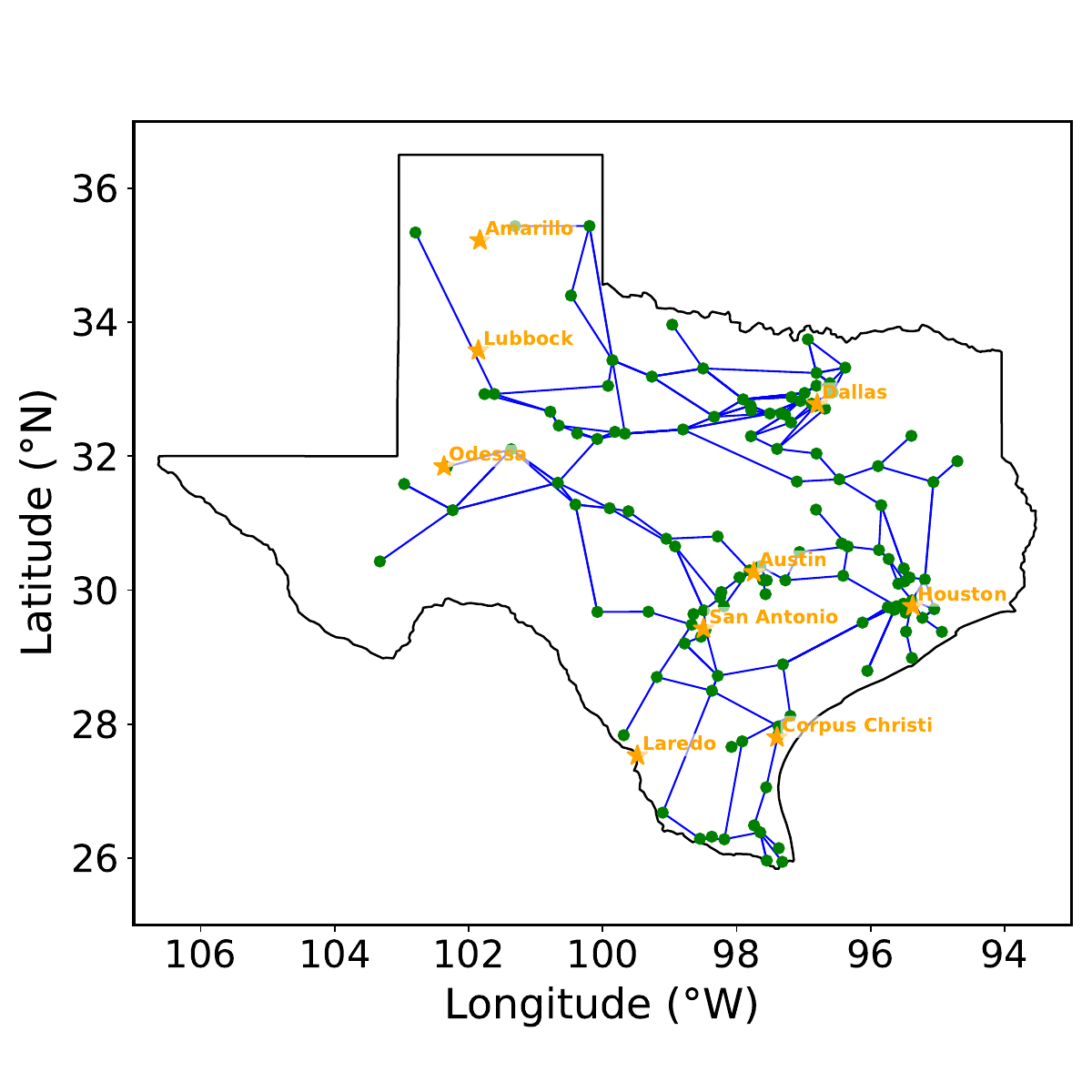}
    \caption{The reference synthetic power grid system. Green dots represent buses and blue lines represent transmission lines.}
    \label{fig:TX-123BT}
\end{figure}

The synthetic grid adapted from \cite{lu2025synthetic} is presented in Fig.~\ref{fig:TX-123BT}. The grid has $N=123$ buses, six generation technologies including both thermal (coal, natural gas, nuclear) and renewable (wind, solar, and hydro), 255 lines of 345 kV high voltage transmission lines assuming uni-direction operation, and no initial storage. The original grid is modeled to reflect the ERCOT grid profile in 2019.  We modify this grid for our case study as follows: First, in the model developed in Section~\ref{sec: math model}, it is assumed that each bus has at most one generator for each generation type. In the original data, a bus can have multiple generators with the same generation technology. For each bus we aggregate all such generators into a single one whose capacity is equal to the summation of the individual generators for each generation technology. We assume that the transmission lines are bi-directional and aggregate the transmission lines connecting two buses $n$ and $n'$, i.e., the capacity of a line is equal to the summation of the capacities of all the lines connecting bus $n$ and $n'$. The resulting grid has 151 generators, 173 transmission lines, and initially zero storage capacity. 

Finally, to account for the spatial availability of renewable generator capacity factor, we calculate the capacity factors for solar and wind as follows
\begin{equation}
\label{eq: capacity factor calculation}
\begin{split}
    & F^{\rm{RN}}_{nitdh} = \frac{\hat{p}^{\rm{gen}}_{nidh}}{c^{\rm{gen},0}_{ni}}
    \ \ \forall n \in \mathcal{N}, i \in \mathcal{I}_{\rm{RN}}(n), d \in \mathcal{D}, h \in \mathcal{H},
\end{split}
\end{equation}
where $\hat{p}^{\rm{gen}}_{nidh}$ is the power generated from technology $i$ at node $n$ in day $d$ and hour $h$ as provided in the original data \cite{lu2025synthetic}. For hydro generation, we assume the capacity factor remains constant across hours, days, and years, and is equal to 41\% \cite{mirletz2024annual}. 

We assume that the construction time for nuclear plants is six years \cite{mirletz2024annual}, for natural gas, hydropower, wind, and transmission lines three years \cite{mirletz2024annual, IEA2023_leadtime}, for coal plants five years \cite{mirletz2024annual}, and for solar and storage one year \cite{mirletz2024annual}. Finally, the round trip efficiency of the storage ($\eta^{\rm{RTE}}$) is set to 85\% \cite{mirletz2024annual}, and the charging and discharging efficiency rate are assumed to be the same, i.e., $\eta^{\rm{charge}} = \eta^{\rm{disch}} = \sqrt{\eta^{\rm{RTE}}}$, and the installed batteries have four hours of storage duration $(H^{\rm{stor}}=4)$.

\begin{figure}[h]
  \centering
  \includegraphics[scale=0.55]{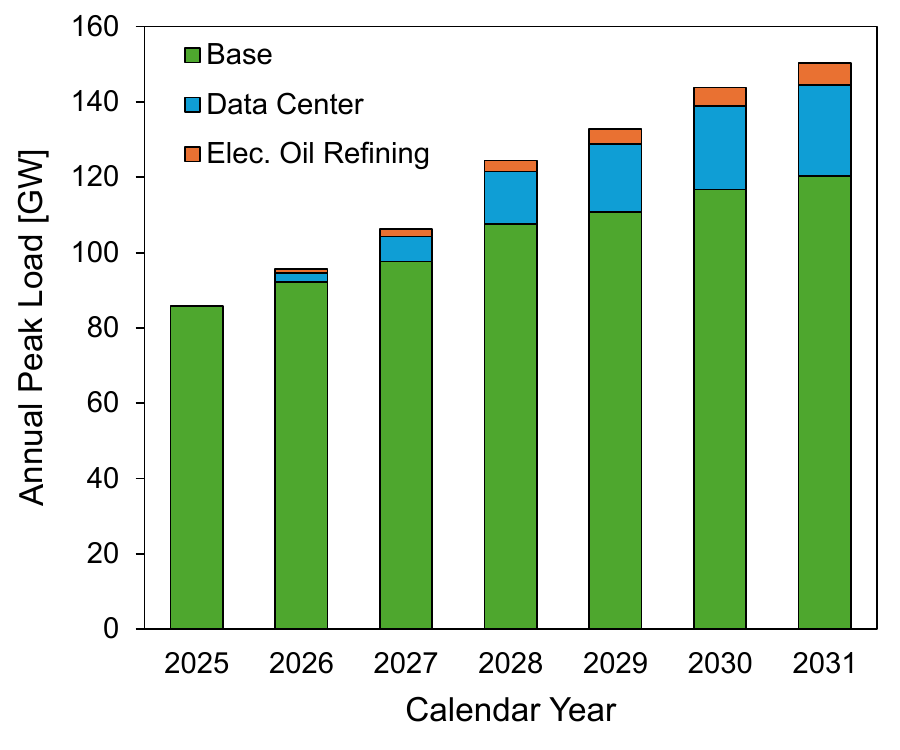}
  \caption{Annual peak load growth scenario breakdown by demand sources over the seven-year planning horizon (2025-2031).}
  \label{fig: loadGrowthScenario}
\end{figure}

\subsection{Input parameters}\label{sec: model parameters}
Operational and economic parameters are retrieved from multiple sources. Initial grid generation and transmission capacities, hourly base load profile, hourly renewable availability, and transmission lines' reactances are retrieved from \cite{lu2025synthetic}. Capital and operational costs of generation and storage technologies, ramping rates, heat rate of thermal generators, and storage round-trip efficiency are obtained from \cite{mirletz2024annual}. 

Note that above-mentioned cost parameters follow the \textit{R\&D scenario} for financial assumptions and the \textit{Moderate scenario} for technology innovation. Specifically, we adopt the following technology representations: for coal, nuclear, and natural gas, we select Coal - New (without carbon capture), Nuclear - Large, Natural Gas 2-on-1 Combined Cycle (H-frame) without CCS, respectively. For renewable technologies, we use Land-based Wind - Class 4 - Technology 1, Utility PV - Class 4, and Hydropower - NPD 2 for wind, solar, and hydro, respectively. For storage, Utility-Scale Battery Storage - 4Hr is selected. Detailed descriptions of these technologies and associated cost assumptions are provided in \cite{mirletz2024annual}.

Transmission capital costs are obtained from \cite{andrade2016estimation}, while projections of fuel costs for thermal generators (natural gas, coal, and uranium) are taken from \cite{decarolis2023annual}. Demand curtailment cost is set to $\$5000/MWh$ \cite{li2022mixed}, and the generation curtailment cost for all generation technologies is set to $\$100/MWh$. Finally, as defined in Eq.~\ref{eq: total costs}, all monetary values are discounted to their net present value using 2022 as the financial base year ($t_{\rm{base}}$).

In this case study, we assume that new nuclear power plants cannot be built until 2029, since commercial demonstrations of new nuclear reactors are expected to be completed by 2030 \cite{mirletz2024annual}. Regarding the existing nuclear power plants, reference \cite{mirletz2024annual} does not provide operational costs for nuclear generators before 2029. We assume that the fixed (FOM) and variable (VOM) operation and maintenance costs do not change between the beginning of the planning horizon and 2029, and assume that these costs are equal to their 2030 counterparts throughout the planning horizon.

\begin{table*}[htbp]
\centering
\caption{Input parameters and data sources used in the case study. Note that some parameters require unit conversion to ensure unit consistency in the constraint equations and the objective function.}
\label{tab:input_parameters}
\small
\setlength{\tabcolsep}{3pt}
\renewcommand{\arraystretch}{1.1}
\begin{tabularx}{\textwidth}{lllllXl}
\hline
Category & Parameter & Symbol & Unit & Value & Note & Ref. \\
\hline
Reference grid & Initial capacity & $c^{\rm{gen},0}_{ni},\ c^{\rm{trans},0}_{nn'}$ & MW & varies & Generation, transmission & \cite{lu2025synthetic} \\
\quad & Initial base load profile & $D^{\rm{base,0}}_{ntdh}$ & MW & varies & Hourly load profile & \cite{lu2025synthetic} \\
\quad & Renewable capacity factor & $F^{\rm{RN}}_{nitdh}$ & - & varies & Calculated through Eq.~\ref{eq: capacity factor calculation} & \cite{lu2025synthetic} \\
\quad & Transmission line reactance & $X_{nn'}$ & p.u. & varies & For transmission power & \cite{lu2025synthetic} \\

Electricity demand & Annual base demand & $E^{\rm{base}}_{t}$ & MWh & varies & Used for Eq.~\ref{eq: demand base} & \cite{ERCOT2025LTLF} \\
\quad & Data center (DC) peak load & $P^{\rm{DC}}_{t}$ & MW & varies & Used for Eq.~\ref{eq: demand DC} & \cite{ERCOT2025LTLF} \\
\quad & DC load factor & $LF^{\rm{DC}}$ & - & 0.9 & Used for Eq.~\ref{eq: demand DC} & \cite{ERCOT2025LTLF} \\
\quad & DC load spatial distribution & $\psi_{c}$ & - & varies & Used for Eq.~\ref{eq: demand DC} & \cite{NREL_SpeedToPower} \\
\quad & \makecell[l]{Total heat demand \\ (oil refining)} & $Q^{\rm{OR}}$ & GW & 18.97 & Used for Eq.~\ref{eq: demand EOR} & \cite{mcmillan2018industrial} \\
\quad & Electrification ratio & $\phi_{t}$ & \% & +5\%/yr & Used for Eq.~\ref{eq: demand EOR} & - \\
\quad & Joule heating efficiency & $\eta_{\rm{elec}}$ & - & 0.97 & Used for Eq.~\ref{eq: demand EOR} & \cite{zuberi2022electrification} \\
\quad & \makecell[l]{Electrification load\\ spatial distribution} & $\psi_{e}$ & - & varies & Used for Eq.~\ref{eq: demand EOR} & \cite{mcmillan2018industrial} \\

Construction time & Nuclear & $\omega^{\rm{gen}}_{\rm{nuclear}}$ & yr & 6 & New builds & \cite{mirletz2024annual} \\
\quad & Coal & $\omega^{\rm{gen}}_{\rm{coal}}$ & yr & 5 & New builds & \cite{mirletz2024annual} \\
\quad & Natural gas  & $\omega^{\rm{gen}}_{NG}$ & yr & 3 & New builds  & \cite{mirletz2024annual} \\
\quad & Hydro  & $\omega^{\rm{gen}}_{hydro}$ & yr & 3 & New builds  & \cite{mirletz2024annual} \\
\quad & Wind  & $\omega^{\rm{gen}}_{wind}$ & yr & 3 & New builds  & \cite{mirletz2024annual} \\
\quad & Solar & $\omega^{\rm{gen}}_{\rm{solar}}$ & yr & 1 & New builds & \cite{mirletz2024annual} \\
\quad & Storage & $\omega^{\rm{stor}}$ & yr & 1 & New builds & \cite{mirletz2024annual} \\
\quad & Transmission & $\omega^{\rm{trans}}$ & yr & 3 & New builds & \cite{IEA2023_leadtime} \\

Storage operation & Round-trip efficiency & $\eta^{\rm{RTE}}$ & - & 0.85 & For battery storage & \cite{mirletz2024annual} \\
\quad & Charge efficiency & $\eta^{\rm{charge}}$ & - & $\sqrt{0.85}$ & $\eta^{charge}=\sqrt{\eta^{\rm{RTE}}}$ & - \\
\quad & Discharge efficiency & $\eta^{\rm{disch}}$ & - & $\sqrt{0.85}$ & $\eta^{disch}=\sqrt{\eta^{\rm{RTE}}}$ & - \\
\quad & Storage duration & $H^{\rm{stor}}$ & hr & 4 & Charge/Discharge duration & - \\

Capital cost (CAPEX) & Generation & $\alpha^{\rm{gen}}_{it}$ & \$/kW & varies & Unit capital cost of generation technology   & \cite{mirletz2024annual} \\
\quad & Storage & $\alpha^{\rm{stor}}_{t}$ & \$/kW & varies & Unit capital cost of storage & \cite{mirletz2024annual} \\
\quad & Transmission & $\alpha^{\rm{trans}}_{t}$ & \$/MW-mi & 0.93 & Unit capital cost of transmission & \cite{andrade2016estimation} \\

Operational cost (OPEX) & Generation FOM & $\beta^{\rm{gen}}_{it}$ & \$/kW-yr & varies & For all generation technology & \cite{mirletz2024annual} \\
\quad & Generation VOM & $\gamma^{\rm{gen}}_{it}$ & \$/MWh & varies & For thermal units only & \cite{mirletz2024annual} \\
\quad & Fuel cost & $\gamma^{\rm{fuel}}_{it}$ & \$/MMBtu & varies & For thermal units only & \cite{decarolis2023annual} \\
\quad & Heat rate & $HR_{it}$ & MMBtu/MWh & varies & For thermal units only & \cite{mirletz2024annual} \\

Curtailment cost & Demand & $\delta$ & \$/MWh & 5000 & Load shedding penalty & \cite{li2022mixed} \\
\quad & Generation & $\zeta$ & \$/MWh & 100 & For all generation technology & - \\

\hline
\end{tabularx}
\end{table*}

\begin{table*}
\centering
\caption{Modeled load growth scenario: annual electricity demand and peak load over the planning horizon from 2025 to 2031}
\label{tab: loadScenario}
\begin{tabular}{c c r r r r r r r r}
\toprule
\multirow{2}{*}{Year} 
& \multirow{2}{*}{\makecell{Electrification\\Ratio}}
& \multicolumn{4}{c}{Electricity Demand [TWh]} 
& \multicolumn{4}{c}{Peak Load [GW]} \\
\cmidrule(lr){3-6} \cmidrule(lr){7-10}
& & Base & DC & EOR & Total 
& Base & DC & EOR & Total \\
\midrule
2025 & 0\%  & 485.90 & 0.00  & 0.00  & 485.90  & 85.76 & 0.00 & 0.00 & 85.76 \\
2026 & 5\%  & 538.94 & 19.18 & 8.56  & 566.69  & 92.22 & 2.43 & 0.98 & 95.63 \\
2027 & 10\% & 595.63 & 52.51 & 17.13 & 665.27  & 97.63 & 6.66 & 1.96 & 106.25 \\
2028 & 15\% & 685.31 & 109.61 & 25.69 & 820.62 & 107.64 & 13.90 & 2.93 & 124.48 \\
2029 & 20\% & 747.59 & 141.87 & 34.26 & 923.72 & 110.86 & 18.00 & 3.91 & 132.76 \\
2030 & 25\% & 808.97 & 174.83 & 42.82 & 1026.62 & 116.77 & 22.18 & 4.89 & 143.83 \\
2031 & 30\% & 846.99 & 190.75 & 51.39 & 1089.13 & 120.33 & 24.20 & 5.87 & 150.39 \\
\bottomrule
\end{tabular}
\end{table*}

\subsection{Time clustering}\label{sec: time clusteting}
The original dataset accompanying the reference grid~\cite{lu2025synthetic} provides hourly-resolution profiles of base load and renewable capacity factors. To maintain computational tractability of the model, we use representative days to capture the evolution of the base demand and renewable capacity factors. The original data accompanying the grid model~\cite{lu2025synthetic} contain: (1) the hourly base load profile for every bus $n$ in year 2019, (2) hourly solar capacity factors for all buses, and (3) hourly wind capacity factors for all buses. We use the k-means clustering \cite{tavenard2020tslearn} with the Euclidean distance as the similarity metric to identify the representative days. The number of representative days $N_{d}$ is determined to be five through sensitivity analysis using both indices of inertia and Silhouette score. An elbow point appears at $N_{d}=5$, where further increasing the number of clusters results in only minor improvements in inertia and clustering performance. The resulting representative days capture diurnal variation and seasonal trends in base load for all buses in hourly-resolution, and variable and intermittent characteristics of solar and wind generation availability.

\subsection{Spatial mapping between buses and counties}

As described Section~\ref{sec: math model}, the large loads are assumed to be provided at a regional level, which for our case study is county. However, from the reference grid some counties do not have any buses, so the large loads can not be satisfied. Thus, for each type of large load, we reassign the demand from such county to the closest county which has at least one bus. Specifically, each bus is first mapped to the county in which it is physically located using FCC Area and Census Block API \cite{FCC_Census_Block_API}. For each country $\ell$ that does not have a bus, we use the great-circle distance between two counties to identify the closest county $\ell'$ with a bus and reallocate the demand from county $\ell$ to $\ell'$. Finally, if a county does not have data center or electrified manufacturing demand, it is not included in the sets $\mathcal{C}$ or $\mathcal{E}$. Under this spatial mapping method, the number of counties with data centers and electrified manufacturing is $N_{c} = 38$ and $N_{e}=10$, respectively.

\subsection{Electricity Demand}
The original grid provides hourly load profile for every bus which we use to define the initial base load hourly profile $D^{\rm{base},0}_{ntdh}$, corresponding to the load that ERCOT exhibited in 2019 \cite{lu2025synthetic}. The projected annual base demand ($E^{\rm{base}}_{t}$) is obtained from ERCOT's load forecast report \cite{ERCOT2025LTLF}. The data center and electrified manufacturing geographic regions correspond to the counties in Texas. For the data centers, the annual peak load ($P^{\rm{DC}}_{t}$) is obtained from \cite{ERCOT2025LTLF} and the spatial distribution ($\psi_c$) is computed using the county-level share of planned data center capacity in Texas \cite{NREL_SpeedToPower}. For manufacturing electrification, we focus on oil refining considering its significant energy consumption \cite{DOE_ITO_2018_MECS}. We focus on the electrification of conventional boilers ($100 \degree C-400 \degree C$) and process heating ($\geq400\degree C$). We use publicly available data \cite{mcmillan2018industrial} to estimate the total heat demand $Q^{\rm{OR}}$ ($Q^{\rm{M}}=Q^{\rm{OR}}$ in Eq.~\ref{eq: demand EOR}) of oil refineries in Texas, which is approximately equivalent to 18.97 GW, assuming it has not increased since 2015. We assume that the electrification ratio $\phi_{t}$ increases by 5\% annually, starting from 0\% in 2025 and the Joule-heating efficiency ($\eta_{\rm{elec}}$) is set equal to 97\% \cite{zuberi2022electrification}. The spatial distribution of electrified manufacturing ($\psi_e$) is computed using the county-level heat duty share out of total oil refining heat duty in Texas using following reference \cite{mcmillan2018industrial}. Following Eqs.~\ref{eq: demand base}, ~\ref{eq: demand DC}, ~\ref{eq: demand EOR}, the modeled load growth scenario in annual electricity demand and peak load is tabulated in Table~\ref{tab: loadScenario}, and the annual peak load and its breakdown by different demand sources are presented in Fig.~\ref{fig: loadGrowthScenario}. Under this specific load growth scenario, data centers and electrified oil refining account for 17.5\% and 4.7\% of the total annual electricity demand at the end of the planning horizon. All input parameters and data sources used in the case study are summarized in Table~\ref{tab:input_parameters}.

\section{Results}\label{sec: results}
In this section, we present the results for the case study described above. The optimization model has approximately $859 \times 10^{3}$ variables, $1.45 \times 10^{6}$ constraints. The model is solved with Gurobi 12.0.0 \cite{gurobi} on an Intel i7 14700 CPU. The solution time varies across individual instances but is approximately one hour.

\subsection{Total cost and new capacity} \label{sec: results_totalCost}
\begin{figure*}
    \centering
    \begin{subfigure}[b]{0.45\textwidth}
        \centering
        \includegraphics[width=\textwidth]{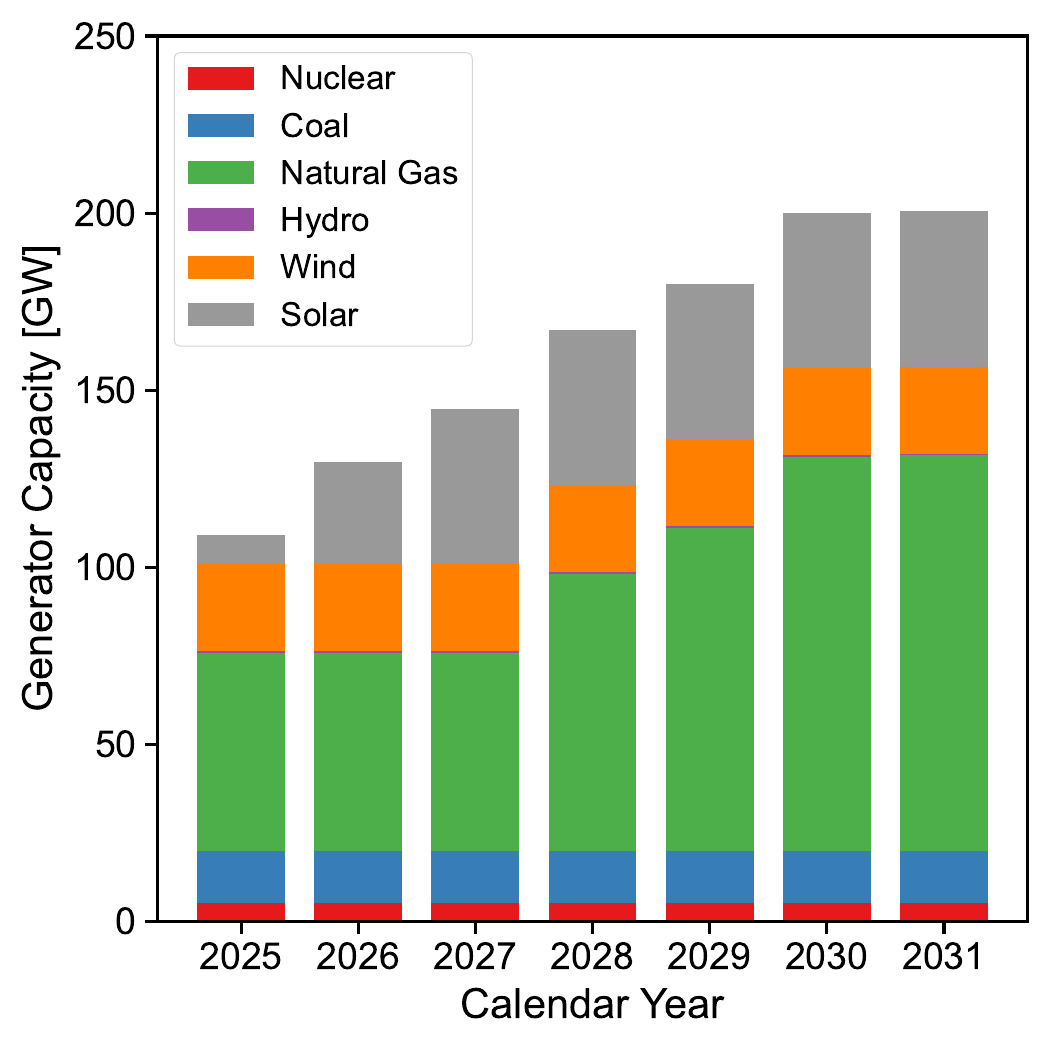}
        \caption{}
        \label{fig: results_generation_cap}
    \end{subfigure}
    \hspace{0.05\textwidth}
    \begin{subfigure}[b]{0.45\textwidth}
        \centering
        \includegraphics[width=\textwidth]{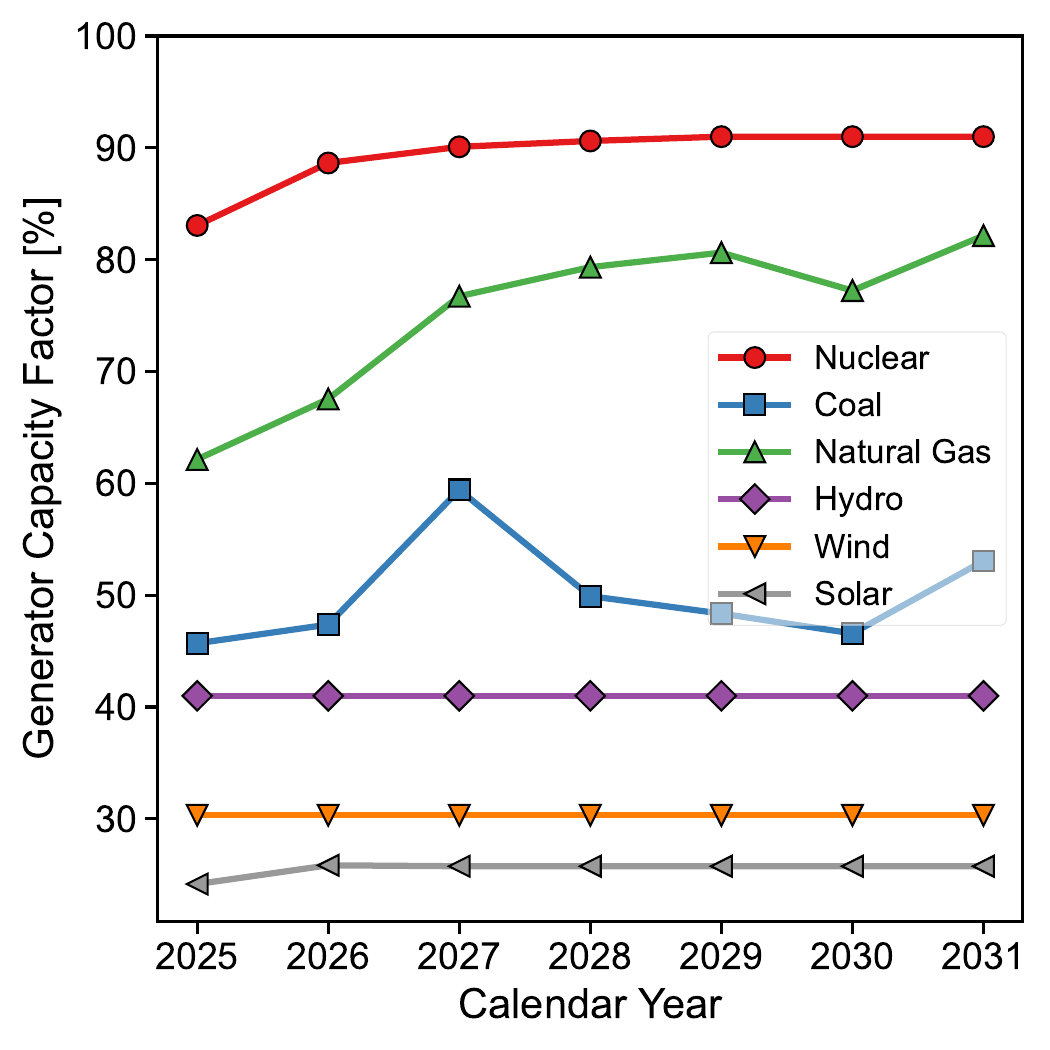}
        \caption{}
        \label{fig: results_generation_CF}
    \end{subfigure}

    \caption{Results of multi-period generation capacity expansion;
    (a) Generator capacity by generation technology;
    (b) Generator capacity factor by generation technology.}
    \label{fig: results_generation}
\end{figure*}

The total cost over the seven-year planning horizon is 319.23 billion USD in 2022 US dollars value. The total generation capacity expands from 109.15 GW in 2025 to 200.47 GW in 2031, an 83.7\% increase over seven years. Natural gas is expanded by 55.7 GW, solar by 35.62 GW, and storage by 19.15 GW. The optimal investment policy in this case study does not include investments in wind, nuclear, or coal. Regarding transmission, the total system-wide capacity increases from 319 GW in 2025 to 322.14 GW in 2031, an increase of approximately 1\%. We note that as described in Section~\ref{sec: referenceGrid} we assumed that the transmission lines are bi-directional, which could affect this transmission expansion decisions. The detailed evolution of the capacity for each generation technology over time is presented in Table~\ref{tab: gridCapResults} and Fig.~\ref{fig: results_generation_cap}.  

\begin{table*}[h]
\centering
\caption{Annual grid generation, transmission, and storage capacity expansion results}
\label{tab: gridCapResults}


\begin{tabular}{l r r r r r r r}
\toprule
\multirow{2}{*}{Capacity [GW]}
& \multicolumn{7}{c}{Calendar Year} \\
\cmidrule(lr){2-8}
 & 2025 & 2026 & 2027 & 2028 & 2029 & 2030 & 2031 \\
\midrule
Nuclear 
& 5.14 & 5.14 & 5.14 & 5.14 & 5.14 & 5.14 & 5.14 \\
Natural gas 
& 56.01 & 56.01 & 56.01 & 78.19 & 91.26 & 111.35 & 111.71 \\
Coal 
& 14.76 & 14.76 & 14.76 & 14.76 & 14.76 & 14.76 & 14.76 \\
Solar   
& 8.26 & 28.88 & 43.89 & 43.89 & 43.89 & 43.89 & 43.89 \\
Wind 
& 24.48 & 24.48 & 24.48 & 24.48 & 24.48 & 24.48 & 24.48 \\
Hydro
& 0.50 & 0.50 & 0.50 & 0.50 & 0.50 & 0.50 & 0.50 \\
Transmission
& 319.00 & 319.00 & 319.00 & 320.64 & 320.92 & 322.06 & 322.14 \\
Storage 
& 0.00 & 19.02 & 19.15 & 19.15 & 19.15 & 19.15 & 19.15 \\
\bottomrule
\end{tabular}
\end{table*}

The difference between generation and transmission capacity expansion depends on the spatial distribution of demand and the grid's initial transmission capacity. Specifically, the majority of the projected data center demand in Texas is located near large cities, such as Dallas, Austin, Houston, and San Antonio, whereas demand from electrified oil refining is located in the coastal area. Thus, new generators are installed in regions with high demand. This result indicates that building new solar and natural gas generators near demand points is more economical than deploying renewable generators in remote areas with higher capacity factors, primarily due to the need for additional transmission expansion. This effect is further reinforced by construction times, as transmission lines take three years to build, compared to one year for solar generation.

\begin{figure*}[t]
    \centering
    \begin{subfigure}[b]{0.45\textwidth}
        \centering
        \includegraphics[width=\textwidth]{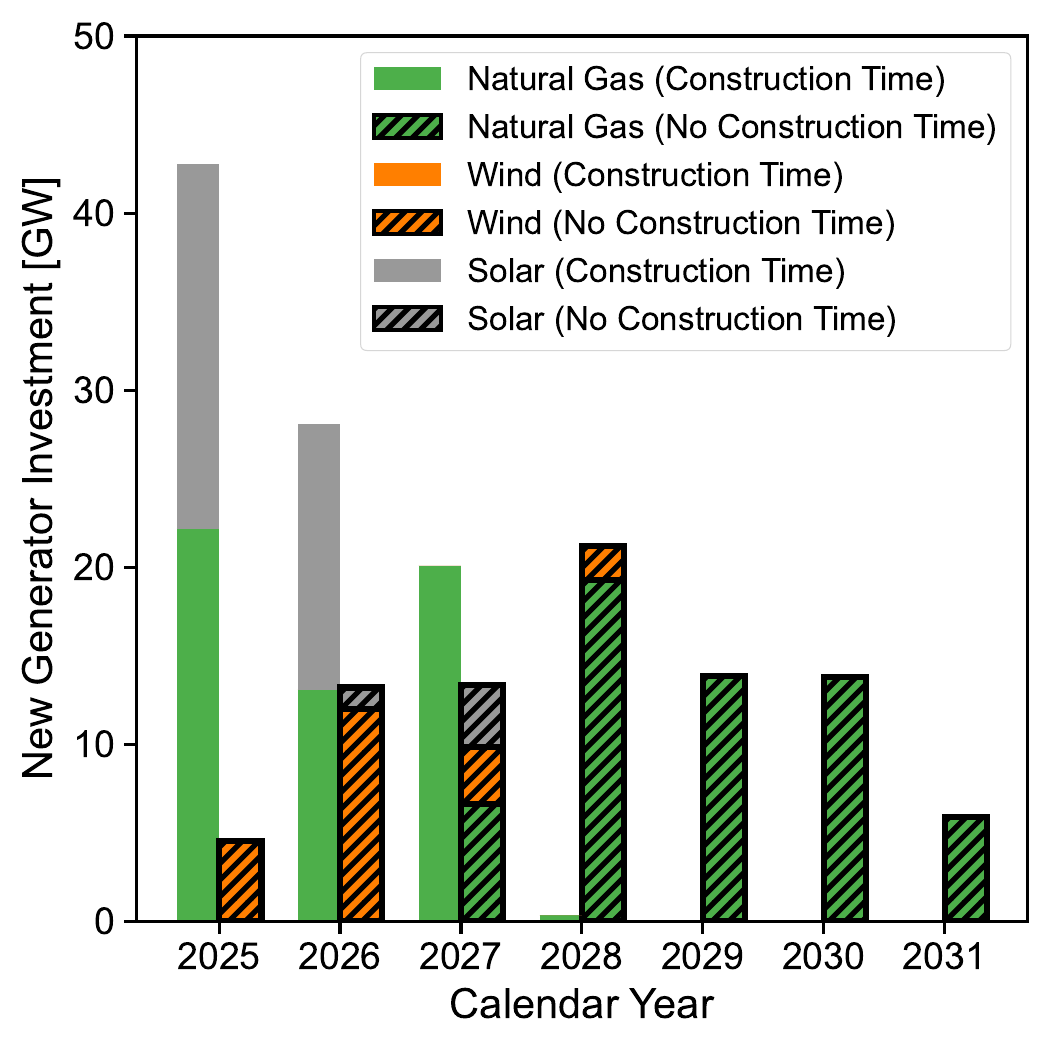}
        \caption{}
        \label{fig: results_const_newGenInv}
    \end{subfigure}
    \hspace{0.05\textwidth}
    \begin{subfigure}[b]{0.45\textwidth}
        \centering
        \includegraphics[width=\textwidth]{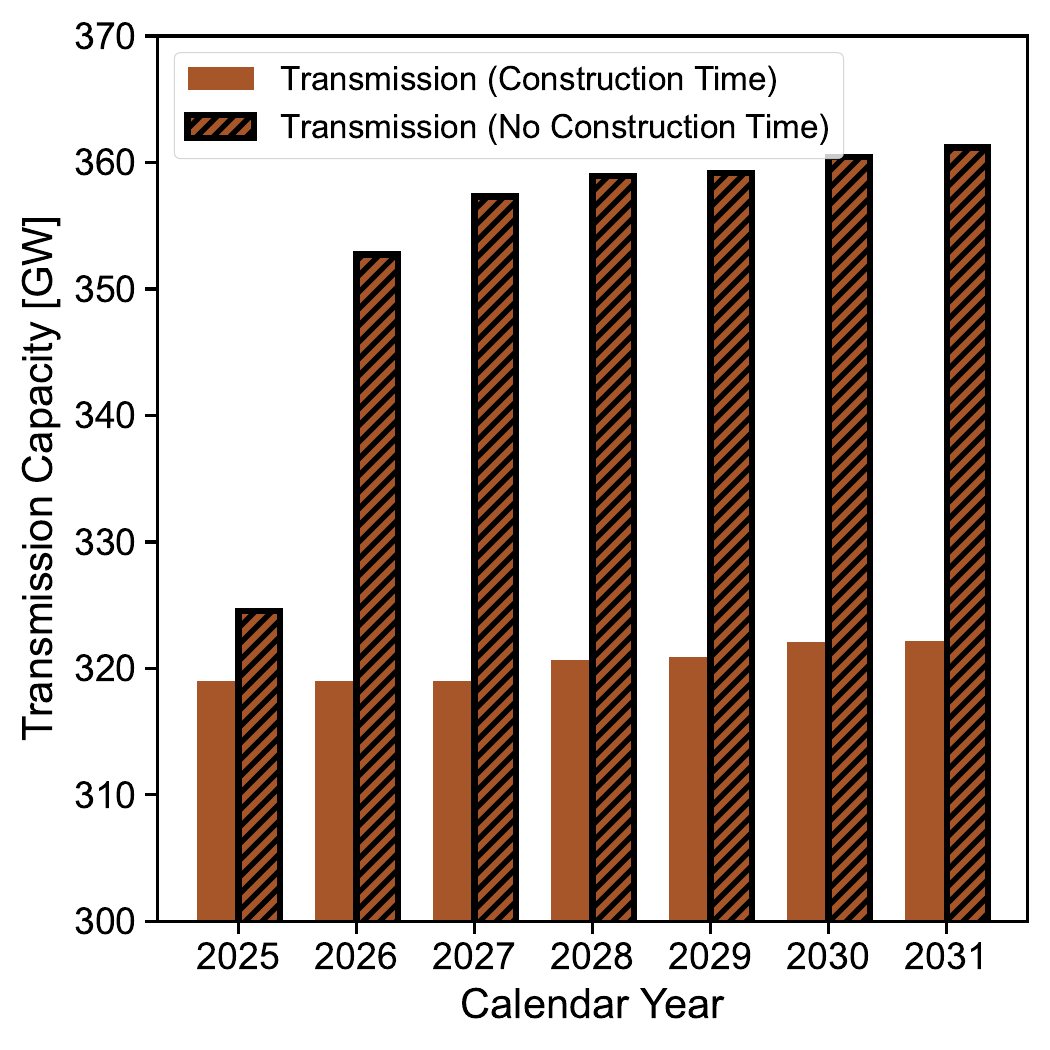}
        \caption{}
        \label{fig: results_const_transCap}
    \end{subfigure}
    \caption{Effects of construction times of assets on capacity expansion results;
    (a) New generator investment;
    (b) Transmission capacity; 
    Solid bars represent the case of considering construction times while hatched bars represent the case of ignoring construction times.}
    \label{fig: results_const}
\end{figure*}

\subsection{The role of solar and thermal generators} \label{sec: results_investmentStrategies}

Analyzing the investment profile further, we observe that the system exploits the different characteristics of renewable and thermal generators. The new generation capacity investments are made in the first three years, leading to a highly front-loaded investment profile. At the beginning of the planning horizon, the increase in demand is met by investing in new solar generation and increasing the capacity factor of existing thermal generators. Specifically, the investment policy exploits the one-year construction time of solar generation to meet the increased demand and minimize demand curtailment. Under this expansion policy, 1.66 TWh of demand is curtailed in the first planning year, which is equivalent to 0.34\% of total annual electricity demand. This implies that new generators with short construction times are critical to mitigate the curtailment penalties.

The construction time of the natural gas facilities creates a time lag in the system. Even though new investments are made in the first planning period, they are not operational until 2028. To reduce demand curtailment and minimize total cost, the utilization of the thermal generators is increased. From Fig.~\ref{fig: results_generation_CF} we observe that for the initial three years, from 2025 to 2027, capacity factors of nuclear, coal, and natural gas are steadily increased close to their maximum levels. These results show that increased utilization of thermal generators is critical for minimizing demand curtailment. We note that for nuclear generation facilities, the capacity factor is at the maximum value for most of the planning horizon. From the fourth planning year, 2028, new natural gas capacity becomes available, and the capacity factor of coal plants declines due to their higher cost. However, as demand grows towards the end of the planning horizon, coal plant utilization increases again to reduce demand curtailment. The capacity factor for solar, wind, and hydro is approximately 25.76\%, 30.32\%, and 41\%, respectively. These results show that the adjustable capacity factors of existing thermal generators reduce costs and can act as a buffer to accommodate lag in generation capacity availability due to construction time.

Finally, the system utilizes natural gas and solar generation to meet the large loads. As described in Section ~\ref{sec: modeling load}, we assume that the demand from large loads is constant each year. Therefore, relying primarily on solar generation would require storage and/or transmission investments to provide steady electricity to large loads throughout the day, especially during periods with low capacity factor, like nighttime. This investment policy would lead to high capital cost for storage. From the results, the capital cost of solar generation is 52.24 billion USD and that of storage is 32.77 billion USD throughout the planning years. On the other hand, natural gas generation can provide steady and dispatchable power to the large load consumers. However, complete reliance would significantly increase demand curtailment due to the construction time lag and increase costs, owing to its higher operating costs relative to solar. Therefore, the optimal investment policy is simultaneously exploiting the short construction time of solar generation and storage and the flexible and high capacity factors of natural gas and other thermal generators.

\subsection{Effect of construction time} \label{sec: results_construction time}

We further analyze the effect of construction time on the investment policy. First, we consider the case in which the construction time for all generation, transmission, and storage assets is zero. For this case, the generation and transmission investments over the planning horizon are shown in Fig.~\ref{fig: results_const}. These results show a fundamentally different investment profile compared to the one that accounts for construction times. First, the investment policy leads to a more distributed generation portfolio consisting of natural gas (59.5 GW), wind (21.66 GW), and solar (4.73 GW), accompanied by substantial expansion of transmission capacity (42.17 GW). Given new wind generators are typically located in remote areas where high capacity factors are available, large-scale transmission investment follows simultaneously, as large loads might not necessarily be located in these areas. At the same time, the system exploits less storage compared to the scenario considering construction time. 


\begin{figure}
    \centering
    \includegraphics[scale=0.5]{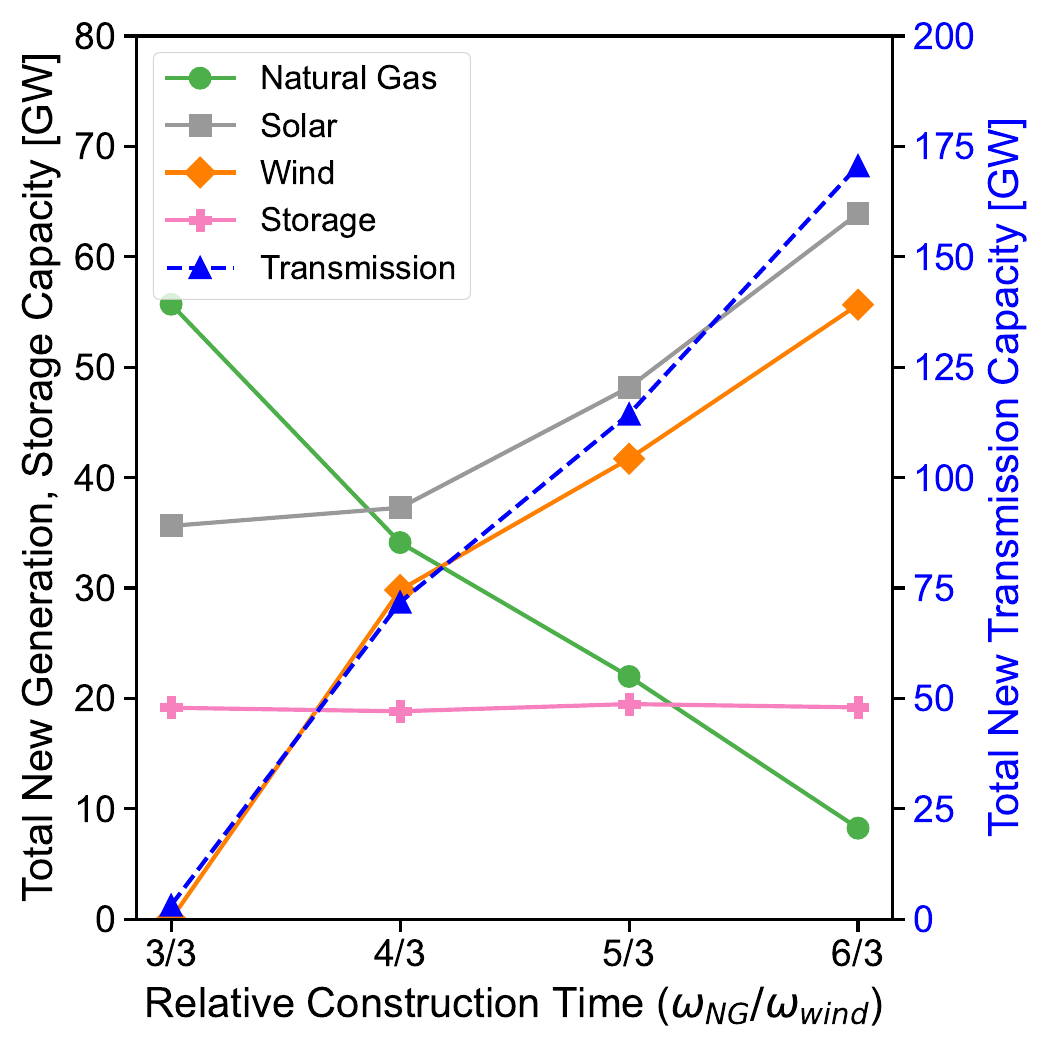}
    \caption{Total new generation, storage, and transmission capacity as a function of the natural gas construction time, assuming the construction times of all other grid assets do not change. 
    Primary y-axis (left) represents total new generation, storage capacity throughout the planning horizon in GW; Auxiliary y-axis (right) represents total new transmission capacity throughout the planning horizon in GW (blue-colored)}
    \label{fig: results_const_NG vs Wind}
\end{figure}

\subsection{Effect of natural gas construction time uncertainty on investment policy}
Finally, we analyze the sensitivity of the optimal investment profile with respect to the construction time of the natural gas generators. We vary the construction time from three to six years in one year increment while fixing the construction times of the others as presented in Section ~\ref{sec: referenceGrid}. For each case, we solve the capacity expansion problem, and the total new capacity of generation, transmission, and storage assets over the entire planning year is presented in Fig.~\ref{fig: results_const_NG vs Wind}. 

The results show that an increase in natural gas construction time significantly shifts investment toward solar and wind and triggers large-scale transmission capacity expansion. The results for $\omega_{\rm{NG}}^{\rm{gen}}/\omega_{\rm{wind}}^{\rm{gen}}=3/3$ correspond to the results presented in Section~\ref{sec: results_totalCost}. In general, the total new capacity of solar and wind increases as the construction time for natural gas power plants increases. The most significant shift occurs when the construction time increases from three to four years, i.e., the construction time of natural gas is greater than that of wind. A further increase in the construction time of natural gas power plants reduces total investments in natural gas and leads to greater deployment of wind and solar generation. The results also highlight the coupling nature between wind and transmission, since locations with high wind generation capacity factors do not necessarily have high demand from large loads. In addition, even if the natural gas construction time is six years, investments in nuclear and coal are not made. Instead, the system relies on renewable generation, transmission and storage to satisfy demand and reduce demand curtailment penalties. We note that investments in storage, in term of total capacity, do not significantly change as the construction time of natural gas increases. Overall, this analysis shows that an increase in natural gas construction time leads to a more geographically distributed energy production system, since the power generation and consumption buses are not necessarily close or even the same. We note that these results are based on the assumption of constant large loads, a fixed spatial distribution, 5000 \$/MW of demand curtailment cost, and a three-year construction time for transmission. 

\section{Conclusions}\label{sec: concl}
Emerging large loads are reshaping the power grid, necessitating rapid expansion over short time horizons. This paper develops a mathematical model to identify optimal grid capacity expansion over a fixed planning horizon such that demand from data centers and electrified manufacturing is satisfied. For a given grid topology, the solution of the model identifies investment strategies and power generation and transmission decisions such that the demand is met and the total cost is minimized. This capacity expansion problem is formulated as a linear optimization problem. We also present a modeling approach to account for the spatial and temporal distribution of large loads from data centers and electrified manufacturing. 

As a case study, the model is used to analyze the expansion of a synthetic grid reflecting ERCOT's network structure and generation and transmission capacity. We consider a seven-year planning horizon from 2025 to 2031 and use publicly available data on expected data center demand. On the manufacturing side, we consider the electrification of process heat and boilers in oil refineries. The demand from both sectors is assumed to change annually; it is constant throughout the year. Under the assumption, the results show that the optimal investment policy expands solar and natural gas generation. Specifically, solar generation, due to its short construction time, can accommodate the initial demand increase and reduce demand curtailment penalties. Simultaneously, the utilization of existing thermal generators is maximized to satisfy the initial demand increase while new natural gas generation facilities become operational due to the construction time lag. Natural gas generation is utilized to meet the steady demand from data centers and electrified manufacturing. The uncertainty regarding the construction time of natural gas has an important effect on the investment policy. Specifically, increasing the construction time of natural gas power plants from three to four years shifts the generation technology mix towards wind and triggers additional transmission capacity expansion. Moreover, even if the construction time is six years, the policy relies on natural gas and renewables, i.e., investments in coal and nuclear are not made. 

Although the proposed framework and case study results provide some insights regarding the effect of large loads on grid expansion, the investment policy depends on several key assumptions made in this work. First, although the spatial distribution of large loads follows given references, the geographical locations of data centers and electrified manufacturing facilities and the extent to which the projected demand will be realized are uncertain. This uncertainty, in combination with the original grid topology, can significantly affect the optimal investment strategy. Second, in the model, the operational behavior of data centers and manufacturing facilities is assumed to be steady throughout the year, reflecting a constant hourly load profile. However, in real-world settings, data centers can flexibly allocate computing demand through task queueing \cite{guo2021integrated, hall2025carbon, wan2025grid}, while manufacturing facilities can adjust their load to some extent through demand response \cite{zhang2016planning}. These operational characteristics and variable dynamics may affect different expansion strategies and investment profiles. Lastly, the model can be readily extended to account for other requirements, such as carbon emissions, power reserves, and renewable penetration. For example, the power generation technology mix could change if carbon emissions are penalized. Finally, modeling extensions can be considered regarding uncertainty arising from construction times, demand realization (magnitude, location, and timing), and technology costs based on technological advancements. We will explore these directions in future work.

\nomenclature[S]{$\mathcal{T}$}{set of planning years}
\nomenclature[S]{$\mathcal{N}$}{set of buses in the grid}
\nomenclature[S]{$\mathcal{I}$}{set of all generation technologies}
\nomenclature[S]{$\mathcal{I}(n)$}{set of generation technologies available at bus $n$}
\nomenclature[S]{$\mathcal{I}_{\rm{TH}}$}{set of thermal generation technologies}
\nomenclature[S]{$\mathcal{I}_{\rm{TH}}(n)$}{set of thermal generation technologies available at bus $n$}
\nomenclature[S]{$\mathcal{I}_{\rm{RN}}$}{set of renewable generation technologies}
\nomenclature[S]{$\mathcal{I}_{\rm{RN}}(n)$}{set of renewable generation technologies available at bus $n$}
\nomenclature[S]{$\mathcal{D}$}{set of representative days}
\nomenclature[S]{$\mathcal{H}$}{set of hours in a day}
\nomenclature[S]{$\mathcal{C}$}{set of geographical regions containing data centers}
\nomenclature[S]{$\mathcal{E}$}{set of geographical regions containing electrified manufacturing}
\nomenclature[S]{$\mathcal{N}(n)$}{set of neighboring buses for bus $n$}
\nomenclature[S]{$\mathcal{N}_{c}$}{set of buses in region $c$}
\nomenclature[S]{$\mathcal{N}_{e}$}{set of buses in region $e$}
\nomenclature[S]{$\mathcal{N}'$}{set of transmission-line pairs $(n,n')$ such that $n' \in \mathcal{N}(n)$ and $n<n'$}

\nomenclature[I]{$n,\ n'$}{indices of buses}
\nomenclature[I]{$i$}{index of generation technologies}
\nomenclature[I]{$t,\ t'$}{indices of planning years}
\nomenclature[I]{$d$}{index of representative days}
\nomenclature[I]{$h$}{index of hours}
\nomenclature[I]{$c$}{index of geographical regions where data centers located}
\nomenclature[I]{$e$}{index of geographical regions where electrified manufacturing facilities located}
\nomenclature[I]{$t_{0}$}{initial year for base demand data}
\nomenclature[I]{$t_{\rm{base}}$}{financial base year}

\nomenclature[V]{$c^{\rm{gen}}_{nit}$}{[MW] new generation capacity at bus $n$ for generation technology $i$ in year $t$}
\nomenclature[V]{$c^{\rm{trans}}_{nn't}$}{[MW] new transmission capacity between buses $n$ and $n'$ in year $t$}
\nomenclature[V]{$c^{\rm{stor}}_{nt}$}{[MW] new storage capacity at bus $n$ in year $t$}

\nomenclature[V]{$C^{\rm{gen}}_{nit}$}{[MW] available generation capacity at bus $n$ for generation technology $i$ in year $t$}
\nomenclature[V]{$C^{\rm{trans}}_{nn't}$}{[MW] available transmission capacity between buses $n$ and $n'$ in year $t$}
\nomenclature[V]{$C^{\rm{stor}}_{nt}$}{[MW] available storage capacity at bus $n$ in year $t$}

\nomenclature[V]{$p^{\rm{gen}}_{nitdh}$}{[MW] power generated at bus $n$ from generation technology $i$ in year $t$, day $d$, and hour $h$}
\nomenclature[V]{$p^{\rm{curt,gen}}_{nitdh}$}{[MW] curtailed generation from technology $i$ at bus $n$ in year $t$, day $d$, and hour $h$}
\nomenclature[V]{$p^{\rm{curt,dem}}_{ntdh}$}{[MW] curtailed demand at bus $n$ in year $t$, day $d$, and hour $h$}
\nomenclature[V]{$p^{\rm{trans}}_{nn'tdh}$}{[MW] power transmitted between buses $n$ and $n'$ in year $t$, day $d$, and hour $h$}
\nomenclature[V]{$p^{\rm{charge}}_{ntdh}$}{[MW] charging power of storage at bus $n$ in year $t$, day $d$, and hour $h$}
\nomenclature[V]{$p^{\rm{disch}}_{ntdh}$}{[MW] discharging power of storage at bus $n$ in year $t$, day $d$, and hour $h$}
\nomenclature[V]{$e^{\rm{stor}}_{ntdh}$}{[MWh] stored energy at bus $n$ in year $t$, day $d$, and hour $h$}
\nomenclature[V]{$p^{\rm{DC}}_{ntdh}$}{[MW] power used to satisfy data center demand at bus $n$ in year $t$, day $d$, and hour $h$}
\nomenclature[V]{$p^{\rm{EOR}}_{ntdh}$}{[MW] power used to satisfy electrified manufacturing demand at bus $n$ in year $t$, day $d$, and hour $h$}
\nomenclature[V]{$\theta_{ntdh}$}{[rad] voltage angle at bus $n$ in year $t$, day $d$, and hour $h$}

\nomenclature[V]{$\Phi^{\rm{CAPEX}}_{t}$}{[\$] capital cost in year $t$}
\nomenclature[V]{$\Phi^{\rm{OPEX}}_{t}$}{[\$] operating cost in year $t$}

\nomenclature[P]{$N$}{total number of buses}
\nomenclature[P]{$N_{\rm{I}}$}{total number of generation technologies}
\nomenclature[P]{$N_{\rm{TH}}$}{number of thermal generation technologies}
\nomenclature[P]{$N_{\rm{RN}}$}{number of renewable generation technologies}
\nomenclature[P]{$N_{\rm{d}}$}{number of representative days}
\nomenclature[P]{$N_{\rm{c}}$}{number of geographical regions containing data centers}
\nomenclature[P]{$N_{\rm{e}}$}{number of geographical regions containing electrified manufacturing}

\nomenclature[P]{$\bar{c}^{\rm{gen}}$}{[MW] upper bound on new generation capacity}
\nomenclature[P]{$\bar{c}^{\rm{trans}}$}{[MW] upper bound on new transmission capacity}
\nomenclature[P]{$\bar{c}^{\rm{stor}}$}{[MW] upper bound on new storage capacity}

\nomenclature[P]{$c^{\rm{gen},0}_{ni}$}{[MW] initial generation capacity at bus $n$ for generation technology $i$}
\nomenclature[P]{$c^{\rm{trans},0}_{nn'}$}{[MW] initial transmission capacity between buses $n$ and $n'$}
\nomenclature[P]{$c^{\rm{stor},0}_{n}$}{[MW] initial storage capacity at bus $n$}

\nomenclature[P]{$\omega^{\rm{gen}}_{i}$}{[year] construction time of generation technology $i$}
\nomenclature[P]{$\omega^{\rm{trans}}$}{[year] construction time of transmission lines}
\nomenclature[P]{$\omega^{\rm{stor}}$}{[year] construction time of storage}

\nomenclature[P]{$P^{\rm{peak}}_{t}$}{[MW] annual peak load of the grid in year $t$}

\nomenclature[P]{$D^{\rm{base}}_{ntdh}$}{[MW] base load at bus $n$ in year $t$, day $d$, and hour $h$}
\nomenclature[P]{$D^{\rm{base},0}_{ndh}$}{[MW] initial base load profile at bus $n$, day $d$, and hour $h$}
\nomenclature[P]{$D^{\rm{DC}}_{ctdh}$}{[MW] data center load in region $c$ in year $t$, day $d$, and hour $h$}
\nomenclature[P]{$D^{\rm{EOR}}_{etdh}$}{[MW] electrified manufacturing load in region $e$ in year $t$, day $d$, and hour $h$}

\nomenclature[P]{$E_{t}$}{[MWh/year] projected total electricity demand in year $t$}
\nomenclature[P]{$E^{\rm{base}}_{t_{0}}$}{[MWh/year] annual base electricity demand in the initial year $t_{0}$}
\nomenclature[P]{$E^{\rm{base}}_{t}$}{[MWh/year] projected annual base electricity demand in year $t$}
\nomenclature[P]{$E^{\rm{DC}}_{t}$}{[MWh/year] annual electricity demand from data centers in year $t$}
\nomenclature[P]{$E^{\rm{EM}}_{t}$}{[MWh/year] annual electricity demand from electrified manufacturing in year $t$}

\nomenclature[P]{$P^{\rm{DC}}_{t}$}{[MW] annual peak load of data center demand in year $t$}
\nomenclature[P]{$LF^{\rm{DC}}$}{load factor of data center}

\nomenclature[P]{$\psi_{c}$}{fraction of total data center demand allocated to region $c$}
\nomenclature[P]{$\psi_{e}$}{fraction of total electrified manufacturing demand allocated to region $e$}
\nomenclature[P]{$\phi_{t}$}{electrification ratio of manufacturing heat demand in year $t$}
\nomenclature[P]{$Q^{\rm{M}}$}{[MW] total heat demand of the manufacturing sector}
\nomenclature[P]{$\eta_{\rm{elec}}$}{joule-heating efficiency}

\nomenclature[P]{$F^{\rm{min}}_{i}$}{minimum capacity factor of thermal generation technology $i$}
\nomenclature[P]{$F^{\rm{max}}_{i}$}{maximum capacity factor of thermal generation technology $i$}
\nomenclature[P]{$F^{\rm{RN}}_{nidh}$}{capacity factor of renewable generation technology $i$ at bus $n$, day $d$, and hour $h$}
\nomenclature[P]{$R^{\rm{ramp}}_{i}$}{[h$^{-1}$] hourly ramping-rate limit of thermal generation technology $i$}

\nomenclature[P]{$S_{\rm{base}}$}{[MVA] base power for the DC-OPF approximation}
\nomenclature[P]{$X_{nn'}$}{[p.u.] reactance of the transmission line between buses $n$ and $n'$}

\nomenclature[P]{$\eta^{\rm{charge}}$}{[-] charging efficiency of storage}
\nomenclature[P]{$\eta^{\rm{disch}}$}{[-] discharging efficiency of storage}
\nomenclature[P]{$H^{\rm{stor}}$}{[h] storage duration}

\nomenclature[P]{$\alpha^{\rm{gen}}_{it}$}{[\$/kW] unit capital cost of generation technology $i$ in year $t$}
\nomenclature[P]{$\alpha^{\rm{trans}}_{t}$}{[\$/MW-mile] unit capital cost of transmission in year $t$}
\nomenclature[P]{$\alpha^{\rm{stor}}_{t}$}{[\$/kW] unit capital cost of storage in year $t$}

\nomenclature[P]{$\beta^{\rm{gen}}_{it}$}{[\$/kW-year] fixed operation and maintenance cost of generation technology $i$ in year $t$}
\nomenclature[P]{$\beta^{\rm{stor}}_{t}$}{[\$/kW-year] fixed operation and maintenance cost of storage in year $t$}
\nomenclature[P]{$\gamma^{\rm{gen}}_{it}$}{[\$/MWh] variable operation and maintenance cost of thermal generation technology $i$ in year $t$}
\nomenclature[P]{$\gamma^{\rm{fuel}}_{it}$}{[\$/MMBtu] fuel cost of thermal generation technology $i$ in year $t$}
\nomenclature[P]{$HR_{it}$}{[MMBtu/MWh] heat rate of thermal generation technology $i$ in year $t$}
\nomenclature[P]{$\hat{\gamma}_{it}$}{[\$/MWh] total variable operating cost of thermal generation technology $i$ in year $t$}

\nomenclature[P]{$\zeta$}{[\$/MWh] curtailment cost of generation}
\nomenclature[P]{$\delta$}{[\$/MWh] curtailment cost of demand}
\nomenclature[P]{$w_{d}$}{weight of representative day $d$}
\nomenclature[P]{$L_{nn'}$}{[mile] distance of the transmission line between buses $n$ and $n'$}
\nomenclature[P]{$Ir$}{interest rate}

\printnomenclature

\section*{Acknowledgments}
Financial support from the Energy Institute at The University of Texas at Austin is gratefully acknowledged. 

\section*{Data Availability}
The code implementation of the model and instructions for reproducing the results are available at the following link:
\url{https://github.com/PSE-Lab/Grid-Capacity-Expansion-under-Data-Center-and-Electrified-Manufacturing-Loads.git}.

\bibliographystyle{elsarticle-num}
\bibliography{references}

\end{document}